\documentclass[10pt, aps, prd, showpacs, twocolumn, notitlepage, nofootinbib, preprintnumbers]{revtex4-1}
\usepackage{amsmath,amssymb,amsfonts,mathrsfs,bbold}
\usepackage{yfonts}
\usepackage{graphicx}
\usepackage[dvipsnames]{xcolor}
\usepackage{hyperref} 
\usepackage{lipsum}
\usepackage{slashed}
\usepackage{simpler-wick}

\usepackage{tikz}
\usetikzlibrary{decorations.pathmorphing, decorations.markings, shapes.geometric, shapes.misc, calc, math}
\tikzstyle{gluon}=[decorate, decoration={coil,aspect=0.8, amplitude=1.5pt,  segment length=3pt}]

\newcommand{\dd}{\ensuremath{\text{d}^2}}

\newcommand{\pvec}{{\underline{p}}}
\newcommand{\qvec}{{\underline{q}}}

\newcommand{\xvec}{{\underline{x}}}

\newcommand{\vecx}{{\ensuremath{\vec{x}}}}
\newcommand{\vecy}{{\ensuremath{\vec{y}}}}

\newcommand{\zv}{{\underline{0}}}

\newcommand{\ul}[1]{\underline{#1}}

\newcommand{\bpsi}{{\bar{\psi}}}

\newcommand{\thalf}{{\tfrac{1}{2}}}

\newcommand{\cald}{{\mathcal{D}}}
\newcommand{\calk}{{\mathcal{K}}}

\newcommand{\calw}{{\mathcal{W}}}
\newcommand{\hcalo}{{\hat{\mathcal{O}}}}

\newcommand{\req}[1]{{Eq.~(\ref{#1})}}

\newcommand{\eik}{{\text{eik}}}
\newcommand{\ag}{{\mathfrak{a}}}

\def\eq#1{{Eq.~(\ref{#1})}}
\def\fig#1{{Fig.~\ref{#1}}}
\newcommand{\ben}{\begin{eqnarray*}}
\newcommand{\een}{\end{eqnarray*}}
\newcommand{\un}[1]{\underline{#1}}

\newcommand{\pd}{\partial}

\newcommand{\as}{\alpha_s}

\begin{document}
\title{Helicity-dependent extension of the McLerran--Venugopalan model} 

\author{Florian Cougoulic} 
         \email[Email: ]{cougoulic.1@osu.edu}
         \affiliation{Department of Physics, The Ohio State
           University, Columbus, OH 43210, USA}
           
\author{Yuri V. Kovchegov} 
         \email[Email: ]{kovchegov.1@osu.edu}
         \affiliation{Department of Physics, The Ohio State
           University, Columbus, OH 43210, USA}

\begin{abstract}
We construct a generalization of the McLerran--Venugopalan (MV) model including helicity effects for a longitudinally polarized target (a proton or a large nucleus). The extended MV model can serve as the initial condition for the helicity generalization of the JIMWLK evolution equation constructed in our previous paper, as well as for calculation of helicity-dependent observables in the quasi-classical approximation to QCD.  
\end{abstract}

\maketitle
\tableofcontents

\section{Introduction}

There is a growing interest among the high-energy QCD community in understanding the sub-eikonal effects and incorporating them into the existing small Bjorken-$x$ formalisms
\cite{Boer:2006rj,Boer:2002ij,Boer:2008ze,Balitsky:2015qba,Altinoluk:2014oxa,Kovchegov:2013cva,Altinoluk:2015gia,Boer:2015pni,Kovchegov:2015pbl,Hatta:2016aoc,Dumitru:2015gaa,Chirilli:2018kkw,Jalilian-Marian:2018iui,Kovchegov:2019rrz,Boussarie:2019icw,McLerran:2018avb,Jalilian-Marian:2019kaf}. Among other things, inclusion of sub-eikonal corrections allows one to access spin dependence of related observables: while in the strict eikonal approximation all the observables are spin-independent, inclusion of at least one sub-eikonal interaction vertex allows one to calculate spin-dependent observables. (Depending on the observable, a sub-eikonal vertex may bring suppression by a power of $x \ll 1$ or may produce a milder suppression by a logarithm of $x$, that is, by $1/\ln (1/x)$, see e.g. \cite{Kovchegov:2012ga,Boer:2015pni,Boussarie:2019vmk}.)

Helicity distribution functions for quarks and gluons in a longitudinally polarized proton along with the orbital angular momentum (OAM) of the proton carried by its quarks and gluons are among the more important and interesting spin-dependent observables. These quantities are essential for our understanding of the proton structure. At the same time, our lack of understanding of the proton spin is manifested by the proton spin puzzle (see \cite{Accardi:2012qut} and references therein): while the helicity sum rules \cite{Jaffe:1989jz,Ji:1996ek,Ji:2012sj} demand that the quark and gluon helicities and OAM should add up to 1/2, the extraction of the parton helicities from the experimental measurements have not yet achieved this number. In addition, little is known experimentally about the parton OAMs. One possible resolution of the proton spin puzzle is that the missing proton spin is carried by the small-$x$ quarks and gluons. Since any given experiment can only probe the values of $x$ down to some small number $x_{min}$, there always remains a possibility that quarks and gluons with $x < x_{min}$ may carry a significant amount of the proton spin undetected by any given (current or future) experiment. It appears that addressing this issue requires progress in our theoretical understanding of helicity and OAM distributions at small $x$. 

An effort to improve our theoretical understanding of small-$x$ helicity and OAM distributions has recently been made in a series of papers \cite{Kovchegov:2015pbl,Kovchegov:2016zex,Kovchegov:2016weo,Kovchegov:2017jxc,Kovchegov:2017lsr,Kovchegov:2018znm,Cougoulic:2019aja,Kovchegov:2019rrz}. The aim of those works was to construct the small-$x$ asymptotics of the quark \cite{Kovchegov:2015pbl,Kovchegov:2016zex,Kovchegov:2016weo,Kovchegov:2017jxc,Kovchegov:2018znm,Cougoulic:2019aja} and gluon \cite{Kovchegov:2017lsr} helicity distributions and OAM \cite{Kovchegov:2019rrz} in perturbative QCD (working at small $x$ but outside of the saturation region). At small $x$ the quark and gluon helicity parton distribution functions (helicity PDFs or hPDFs) can be expressed in terms of the object dubbed the ``polarized dipole scattering amplitude". The latter is an expectation value in the longitudinally polarized target state of a normalized trace of a light-cone Wilson line and a ``polarized Wilson line", which is a linear superposition of light-cone Wilson lines with one and two insertions of sub-eikonal operators depending on helicity \cite{Kovchegov:2017lsr,Kovchegov:2018znm} (one gluon field operator and two quark field operators). Small-$x$ evolution equations resumming all powers of $\as \, \ln^2 (1/x)$ for the polarized dipole amplitude, that the quark hPDF depends on, were written down in \cite{Kovchegov:2015pbl}. Closed equations were obtained in the large-$N_c$ and large-$N_c \& N_f$ limits, with $N_c$ the number of quark colors and $N_f$ the number of flavors. (Compare this with the Balitsky--Kovchegov (BK) equation \cite{Balitsky:1995ub,Balitsky:1998ya,Kovchegov:1999yj,Kovchegov:1999ua} which is a closed equation for the unpolarized scattering derived in the large-$N_c$ limit.) The  large-$N_c$ equations were solved in \cite{Kovchegov:2016weo,Kovchegov:2017jxc}. Similar program was carried out for the gluon hPDF in \cite{Kovchegov:2017lsr}. The large-$N_c \& N_f$ equations have been recently solved numerically in \cite{Kovchegov:2020hgb}.   

To go beyond the large-$N_c$ and large-$N_c \& N_f$ limits, and to automate the way of evolving different correlators made of the standard and ``polarized" Wilson lines, a helicity generalization of Jalilian-Marian--Iancu--McLerran--Weigert--Leonidov--Kovner
(JIMWLK)
\cite{Jalilian-Marian:1997dw,Jalilian-Marian:1997gr,Weigert:2000gi,Iancu:2001ad,Iancu:2000hn,Ferreiro:2001qy} evolution was constructed in our previous paper on the subject \cite{Cougoulic:2019aja} for the flavor-singlet channel. Similar to the original JIMWLK equation, the helicity-JIMWLK (hJIMWLK) equation derived in \cite{Cougoulic:2019aja} is a functional integro-differential equation for the weight functional $\calw [\alpha, \beta, \psi, \bpsi]$. One of the differences between JIMWLK and hJIMWLK is that in the former the weight functional only depends on the eikonal gluon field $\alpha (x) \equiv A^+ (x)$, while the weight functional in the latter also depends on the sub-eikonal gluon field strength [$\beta (x) \equiv F_{12} (x)$] and quark fields ($\psi, \bpsi$) of the target. (Our light-cone coordinates are defined as $x^\pm = (x^0 \pm x^3)/\sqrt{2}$, while Latin indices $i, j =1,2$ denote the components of the transverse vectors ${\un x} = (x^1, x^2)$ with $x_\perp = |{\un x}|$. Our projectile proton is moving in the light-cone plus direction, and the derivation of hJIMWLK in \cite{Cougoulic:2019aja} was carried out in the $\pd_\mu A^\mu =0$ and $A^- =0$ gauges.)

The hJIMWLK equation reads \cite{Cougoulic:2019aja}
\begin{align}
\label{EQ:Helicity_evolution_equation}
{\cal W}_\tau [\alpha,\beta,\psi,\bpsi] = &\ \calw^{(0)} [\alpha,\beta,\psi,\bpsi]  \\ 
& + \int d^3 \tau' \ \calk_h[\tau,\tau'] \cdot {\cal W}_{\tau'}  [\alpha,\beta,\psi,\bpsi], \nonumber
\end{align}
with its integral kernel $\calk_h[\tau,\tau']$ given explicitly in Eq.~(57) of \cite{Cougoulic:2019aja}. Another important distinction between JIMWLK and hJIMWLK is that the evolution of the weight functional in the latter is not just in Bjorken $x$, or, equivalently, rapidity $Y = \ln (1/x)$. In fact, the evolution equation \eqref{EQ:Helicity_evolution_equation} involves the following aggregated variables:
\begin{align}\label{tau_def}
\tau \equiv \{ z, z\, X^2_\perp, z\, Y_\perp^2\}, \ \ \ \tau' \equiv \{ z', z'\, X'^2_\perp, z'\, Y'^2_\perp \}. 
\end{align}
Here $z$ or $z'$ is the longitudinal momentum fraction of the projectile (probe) minus light-cone momentum carried by the partons in question \cite{Kovchegov:2015pbl,Kovchegov:2016zex,Kovchegov:2016weo,Kovchegov:2017jxc,Kovchegov:2018znm,Cougoulic:2019aja}. The lifetimes of the partons in the $x^-$ direction are proportional to $z\, X_\perp^2$ and $z\, Y_\perp^2$ to the left and right of the target proton respectively, with $X_\perp$ and $Y_\perp$ the relevant transverse distances \cite{Cougoulic:2019aja}. The integration measure in \eq{EQ:Helicity_evolution_equation} is $d^3\tau' \equiv \frac{d z'}{z'} \, \dd X'_\perp \dd Y'_\perp$. (Note that in DLA the integrals over the angles of ${\un X}'$ and ${\un Y}'$ are trivial: this is why we write $d^3\tau'$ instead of $d^5\tau'$). The evolution in $\tau$ (instead of rapidity $Y \sim \ln z$) results in the leading-order (LO) hJIMWLK evolution \eqref{EQ:Helicity_evolution_equation} resumming powers of $\as \, \ln^2 (1/x)$ instead of powers of $\as \, \ln (1/x)$ summed up by the standard LO JIMWLK evolution. While the small-$x$ helicity distributions are suppressed by a power of $x$ compared to the unpolarized PDFs, their small-$x$ evolution sums powers of a larger parameter $\as \, \ln^2 (1/x)$ at the leading order \cite{Bartels:1995iu,Bartels:1996wc,Kovchegov:2015pbl,Kovchegov:2016weo,Kovchegov:2016zex,Kovchegov:2017jxc,Kovchegov:2017lsr,Kovchegov:2018znm} (see also \cite{Kirschner:1983di, Kirschner:1985cb,
  Kirschner:1994vc,Kirschner:1994rq,Griffiths:1999dj,Itakura:2003jp,Bartels:2003dj} for details on how this resummation parameter arises for other observables). 

While the kernel of hJIMWLK equation \eqref{EQ:Helicity_evolution_equation} was derived in \cite{Cougoulic:2019aja}, the inhomogeneous term (the initial condition) $\calw^{(0)}$ was not found explicitly in that paper. The goal of the present paper is to construct the functional $\calw^{(0)}$, which is needed for finding the solution of the flavor-singlet hJIMWLK equation.\footnote{Note that in this paper we construct the inhomogeneous term $\calw^{(0)}$ for a large nucleus with the atomic number $A \gg 1$. However, resolution of the proton spin puzzle requires the knowledge of $\calw^{(0)}$ for a proton. Inspired by the success of the unpolarized proton phenomenology at small-$x$ that employs the MV model as the initial condition for BK/JIMWLK evolution (see e.g. \cite{Albacete:2013tpa,Lappi:2013pya}, and references therein), we hope that the helicity-dependent extension of the MV model can also be applied to a proton. The success of the MV model in the unpolarized proton phenomenology is hard to justify theoretically, in absence of the large parameter $A$. It can be that at moderately-small $x$ the parton occupation number in the proton becomes large due to Dokshitzer-Gribov-Lipatov-Altarelli-Parisi (DGLAP) evolution \cite{Gribov:1972ri,Altarelli:1977zs,Dokshitzer:1977sg}, thus diminishing the role of correlations and justifying the use of the MV model. It may also be that the MV model, which combines the lowest-order in $\as$ (often Born-level) contribution to an observable with saturation corrections, captures a sufficient amount of the right physics for many observables to ``work" outside of its original domain of applicability (a large nucleus) and apply to the description of the proton before small-$x$ evolution.}
In the case of the standard JIMWLK evolution, the initial condition is given by the weight functional of the McLerran--Venugopalan (MV) model \cite{McLerran:1993ni,McLerran:1993ka,McLerran:1994vd}, which is Gaussian in the field $\alpha$ \cite{Kovchegov:1996ty}.

Below we construct the functional $\calw^{(0)}$ as a function of $\alpha, \beta, \psi,\bpsi$. The paper is structured as follows. In Section \ref{sec:MV} we rederive the Gaussian functional of the MV model and in the process set up the formalism for the calculation to follow. In Sec.~\ref{sec:Gen_hel} we introduce the necessary sub-eikonal tools and generalize the MV model weight functional to include helicity dependence. The final result for $\calw^{(0)}$ is given in \eq{W_final}, providing the inhomogeneous term for \eq{EQ:Helicity_evolution_equation}. We explicitly verify that it satisfies all the conditions assumed about it in \cite{Cougoulic:2019aja}.


\section{Notations and the McLerran--Venugopalan model}
\label{sec:MV}


\subsection{Nuclear states and averaging procedure}

To begin, let us set up our formalism by reproducing the original McLerran--Venugopalan (MV) model \cite{McLerran:1993ni,McLerran:1993ka,McLerran:1994vd}. To do so, let us replace the target proton by a large nucleus with the atomic number $A \gg 1$. The nucleus is moving ultrarelativistically along the light-cone plus direction. It contains $A$ different nucleons which are independent from each other. The main premise of the MV model is that the small-$x$ gluon field of this large nucleus is given by the solution of the classical Yang-Mills equations, with the source current given by the larger-$x$ ``valence" quarks and gluons in the nucleons of the nucleus (see \cite{Iancu:2003xm,Weigert:2005us,JalilianMarian:2005jf,Gelis:2010nm,Albacete:2014fwa,Kovchegov:2012mbw} for pedagogical presentations of the MV model). These large-$x$ ``valence" quarks and gluons in the nucleons are assumed to be recoilless classical sources of the gluon field. The observables in the MV model are expectation values of operators dependent on this classical gluon field or on the color charge density of the source. 

Let us denote the state of the $i$th nucleon by $| \un{p}_i, p_i^+ \rangle$ normalized as 
\begin{align}\label{state_norm}
\langle p^+, \un{p} | \un{p}', p'^+ \rangle = 2 p^+ (2 \pi)^3 \delta^2(\un{p} - \un{p}') \delta (p^+ - p'^+). 
\end{align}
Employing unit operators
\begin{align}
\mathbb{1} = \prod_{i=1}^A \int \frac{d^2 p_i \, d p_i^+}{(2 \pi)^3 \, 2 p^+} \, | \un{p}_i, p_i^+ \rangle \, \langle p_i^+ , \un{p}_i |
\end{align}
the expectation value of an operator $\hat {\cal O}$ in the nuclear state $| A \rangle$ can be written as \cite{Wu:2017rry,Kovchegov:2013cva,Kovchegov:2015zha},
\begin{align}\label{eq:DMA3}
 &  \langle A | \hat {\cal O} | A \rangle =  \int \prod_{i=1}^A 
  \frac{d^2 P_i \, dP_i^+}{(2\pi)^3} \ d^2 B_i \, d B_i^-
  \\ & \times W \! \left( P_1, \ldots , P_A;  B_1, \ldots , B_A \right) {\cal O} \! \left(
    P_1, \ldots , P_A;  B_1, \ldots , B_A \right) ,  \notag
\end{align}
in terms of the nucleon Wigner distribution
\begin{align}\label{Wigner}
&   W \left( P_1, \ldots , P_A;  B_1, \ldots , B_A
  \right) \\ & = \int \prod_{i=1}^A \frac{d^2 (p_i - p'_i) \, d(p_i^+ - p_i^{\prime
      +})}{(2\pi)^3} \frac{2 P_i^+}{2 p_i^+ \, 2 p_i^{\prime
      +}}  \notag \\ & \times e^{- i (p_i^+ -
    p_i^{\prime +}) B_i^- + i (\un{p}_i - \un{p}'_i) \cdot
    \un{B}_i} \prod_{j=1}^A \langle p_j^+ , \un{p}_j | A
  \rangle \prod_{k=1}^A \langle A | \un{p}'_k, p_k^{\prime +} \rangle  \notag
\end{align}
with $P_i = (p_i + p'_i)/2$ and the operator's expectation value in the nucleon states
\begin{align}
 &  \! \!   {\cal O} \left( P_1, \ldots , P_A;  B_1, \ldots , B_A \right) \! = \! \! \int \! \prod_{i=1}^{A} \!
  \frac{d^2 (p_i - p'_i) d(p_i^+ - p_i^{\prime +})}{(2 \pi)^3 \, 2 P_i^+} \notag \\  & \times e^{i (p_i^+
    - p_i^{\prime +}) B_i^- - i (\un{p}_i - \un{p}'_i) \cdot
    \un{B}_i} \! \prod_{j=1}^A  \langle p_j^{\prime +} , \un{p}'_j | \hat {\cal O} \! \prod_{k=1}^A | \un{p}_k, p_k^+ \rangle .
\end{align}
Here $B_i = (B_i^-, \un{B}_i)$ specify the trajectories of the (centers of the) nucleons. In a large nucleus with many nucleons, $A \gg 1$, the nucleons can be assumed to be independent from each other at the leading order in $A$. This leads to the factorization of the Wigner distribution into a product of individual nucleon's Wigner distributions, 
\begin{align}\label{Wfact}
& W \left( P_1, \ldots , P_A;  B_1, \ldots , B_A \right) = \frac{1}{A!} \\ & \times \, \left[ W (P_1,  B_1) \ldots W (P_A , B_A) + \mbox{permutations} \right] , \notag
\end{align}
where by `permutations' we mean different pairings of $P_i$ and $B_j$ in the arguments of the Wigner functions. In the MV model, neglecting the possible orbital motion of the nucleons in the nucleus \cite{Kovchegov:2013cva}, the classical nucleon Wigner distribution is approximated by\footnote{OAM in the Wigner distribution will manifest itself as a non-zero $\pvec$ for the ``valence" quarks and gluons  \cite{Kovchegov:2013cva}, which would translate into terms proportional to $z\pvec$ in the small-$x$ quark or gluon field, with $z=k^+/p^+ \ll 1$ the fraction of the source's longitudinal momentum carried by the field.
Unpolarized or longitudinally-polarized target nucleus has no preferred transverse direction after the impact parameter integral ${\un b}$ is carried out: hence, only even powers of $z\pvec$ are non-vanishing for unpolarized and helicity observables, with the leading order-$\sim z^2 \pvec^2$ term being sub-sub-eikonal, and thus negligible even with the precision of our sub-eikonal helicity calculations below. In addition, the even powers of $z\pvec$ are independent of the sign of $\pvec$ (i.e., of the sign of OAM), and, therefore, can be safely neglected in applications of the MV model to helicity calculations.}
\begin{align}
  \label{eq:WigMV}
  W_{cl} \left( p, b \right) = \frac{1}{A} \, \rho_A (b^-, \un{b}) \,
  (2 \pi)^3 \, \delta \left( p^+ - \frac{P^+}{A} \right) \, \delta^2
  (\un{p})
\end{align}
where $\rho_A (b^-, \un{b})$ is the nucleon number density normalized such
that
\begin{align}
  \label{eq:density}
  \int d^2b \, d b^- \, \rho_A (b^-, \un{b}) = A 
\end{align}
and $P^+$ is the light-cone momentum of the entire nucleus. Note that our Wigner distribution is normalized as 
\begin{align}
\int \prod_{i=1}^A 
  \frac{d^2 P_i \, dP_i^+}{(2\pi)^3} & d^2 B_i \, d B_i^- \, W \! \left( P_1, \ldots , P_A;  B_1, \ldots , B_A \right) \notag \\ & = \langle A | A \rangle = 1. 
\end{align}

Substituting Eqs.~\eqref{Wfact}
and \eqref{eq:WigMV} into \eq{eq:DMA3} we arrive at
\begin{align}
  \label{eq:DMA4}
  \langle A | \hat {\cal O} | A \rangle = 
  \Bigg[ \int \prod_{i=1}^{A} d^2 B_i \, d B_i^- \frac{1}{A} \, \rho_A (B_i^-,
  \un{B}_i) \Bigg] \\ \times \, {\cal O} \left( B_1, \ldots , B_A \right). \notag
\end{align}
Here we have suppressed the dependence on nucleons momenta in ${\cal O} \left( B_1, \ldots , B_A \right)$, since the momentum of the nucleus is equally distributed among all the nucleons (see \eq{eq:WigMV}) resulting in each nucleon carrying the same momentum $P_i = (P^+/A, \un{0})$. Defining $\Delta_i = p'_i - p_i$ we can use this equal-momentum distribution to write (cf. Appendix~A of \cite{Kovchegov:2019rrz})
\begin{align}\label{Odef}
& {\cal O} \left( B_1, \ldots , B_A \right) \! = \! \! \int \! \prod_{i=1}^{A}
  \frac{d^2 \Delta_i  d \Delta_i^+ }{(2 \pi)^3 \, 2 (P^+/A)} \, e^{-i \Delta_i^+ \, B_i^- + i \un{\Delta}_i \cdot
    \un{B}_i}  \\  & \times \prod_{j=1}^A  \left\langle \frac{P^+}{A} + \thalf \Delta_j^{+} , \thalf \un{\Delta}_j \right| \hat {\cal O} \prod_{k=1}^A \left| -\thalf \un{\Delta}_k, \frac{P^+}{A} - \thalf \Delta_k^+ \right\rangle . \notag 
\end{align}

The expression \eqref{eq:DMA4} implies that averaging in the nuclear state in the MV model is done by averaging over the positions of the nucleons in the nucleus, in agreement with the standard procedure \cite{Kovchegov:1996ty,Kovchegov:1997pc,Kovchegov:1997ke,Kovchegov:2012mbw}. The averaging in each nucleon's state is done assuming that the large-$x$  ``valence" quarks and gluons serve as sources for the classical gluon field: hence, the averaging is done over such ``classical" partonic sources \cite{McLerran:1993ni,McLerran:1993ka,McLerran:1994vd,Kovchegov:1996ty,JalilianMarian:1996xn,Kovchegov:1997pc,Kovchegov:1997ke,Kovchegov:2012mbw}. One ends up with the ``doubling" for formula \eqref{eq:DMA4}:
\begin{align}
  \label{eq:DMA5}
  & \langle A | \hat {\cal O} | A \rangle = 
  \Bigg[ \int \prod_{i=1}^{A} d^2 B_i \, d B_i^- \frac{1}{A} \, \rho_A (B_i^-,
  \un{B}_i) \Bigg] \\ & \times \Bigg[ \int \prod_{j=1}^{N_i} d^2 b_{ij} \, d b_{ij}^- \frac{1}{N_i} \, \rho_{N_i} (b_{ij}^- - B_i^-,
  \un{b}_{ij} -  \un{B}_i) \Bigg] \notag \\ & \times \, {\cal O} \left( b_{11}, \ldots , b_{A,N_A} \right). \notag
\end{align}
Here we assume that the $i$th nucleon has $N_i$ ``valence" (large-$x$) partons at positions $b_{i1}, \ldots , b_{i N_i}$, which are distributed inside the nucleon with the number density $\rho_{N_i}$ normalized similar to \eqref{eq:density}, 
\begin{align}
  \label{eq:densityN}
  \int d^2b \, d b^- \, \rho_{N_i} (b^-, \un{b}) = N_i .
\end{align}
We will also assume that the typical distance between the partons in a nucleon (in the nuclear rest frame) is of the order of the size of the nucleon, $r_p \sim 1/\Lambda_{QCD}$ with $\Lambda_{QCD}$ the non-perturbative QCD scale \cite{Kovchegov:1996ty}: this means that the large-$x$ parton density in the nucleon is not very high. The operator average ${\cal O} \left(b_{11}, \ldots , b_{A,N_A} \right)$ is calculated similar to \eq{Odef}, but now with the parton states instead of the nucleon ones:  
\begin{align}\label{Odef2}
& {\cal O} \left( b_{11}, \ldots , b_{A,N_A} \right) \! = \! \! \int \! \prod_{i=1}^{A} \prod_{j=1}^{N_i} 
  \frac{d^2 \Delta_{ij}  d \Delta_{ij}^+ }{(2 \pi)^3 \, 2 P_{ij}^+} \, e^{-i \Delta_{ij}^+ \, b_{ij}^- + i \un{\Delta}_{ij} \cdot
    \un{b}_{ij}} \notag \\  & \times \prod_{i'=1}^A  \prod_{j'=1}^{{N_{i'}}} \left\langle P_{i'j'}^+ + \thalf \Delta_{i'j'}^{+} , \thalf \un{\Delta}_{i'j'} \right| \hat {\cal O} \notag \\  & \times\prod_{i''=1}^A  \prod_{j''=1}^{{N_{i''}}} \left| -\thalf \un{\Delta}_{i''j''}, P_{i''j''}^+ - \thalf \Delta_{i''j''}^+ \right\rangle ,
\end{align} 
where $P_{ij}^+$ denotes the momentum of the parton $j$ in nucleon $i$. Note that expectation value in a nucleon state implies averaging over parton colors and polarizations, which are not shown explicitly in \eq{Odef2}.


\subsection{Charge density}

The 3-dimensional color charge density $\rho^a$ in covariant (Feynman) $\pd_\mu A^\mu =0$ gauge is usually defined for a given configuration of nucleons and partons in the nucleus as \cite{Kovchegov:1996ty,Kovchegov:1997pc}
\begin{equation}\label{color_dens}
\rho^a_{cov}(x^-, \un{x}) = g \sum_{i=1}^A \sum_{j=1}^{N_i} t_{R_{ij}}^a \, \delta^{2}(\un{x} -\un{b}_{ij}) \, \delta (x^- - b^-_{ij})
\end{equation}
where $t_{R_{ij}}^a$ is the color generator associated to the parton $j$ in the nucleon $i$ in an irreducible representation (irrep) $R_{ij}$ and $g$ is the strong coupling. This expression for the density can be obtained from \eq{Odef2}. Since we need to construct similar densities for the sub-eikonal helicity case below, let us first explicitly demonstrate how \eq{color_dens} results from \eq{Odef2} and which operator has to be used in the latter to obtain the color charge density.  

We begin with the free field operators, which in the light-front perturbation theory are written in terms of the creation and annihilation operators as \cite{Lepage:1980fj,Brodsky:1997de} 
\begin{subequations}\label{ppA}
\begin{align}
\psi^i_0 (x) &= \!\!  \int\limits_{{p},\sigma} 
\left[ \hat{b}^i_{p,\sigma} \, {u}_\sigma(p)e^{-i{p}\cdot{x}} 
+ \hat{d}^{i \, \dagger}_{p,\sigma} \, {v}_\sigma(p) e^{i{p}\cdot{x}} \right], \label{psibd} \\
\bar{\psi}_0^i (x) &= \!\!  \int\limits_{{p},\sigma} 
\left[ \hat{b}^{i \, \dagger}_{p,\sigma}\, \bar{u}_\sigma(p)e^{i{p}\cdot{x}} 
+ \hat{d}^i_{p,\sigma} \, \bar{v}_\sigma(p) e^{-i{p}\cdot{x}} \right], \label{psi_bar_bd}  \\
A^{a, \mu}_0 (x) &= \!\! \int\limits_{{p},\lambda} 
\left[ \hat{a}^a_{p,\lambda} \, {\epsilon}^\mu_\lambda(p)e^{-i{p}\cdot{x}} 
+ \hat{a}^{a \, \dagger}_{p,\lambda} \, \epsilon^{* \, \mu}_\lambda(p) e^{i{p}\cdot{x}} \right], \label{Aaa}
\end{align}
\end{subequations}
for the quark and gluon fields. Here $i$ and $a$ are the fundamental and adjoint color indices, while the quark flavor indices are suppressed. All the fields are taken at $x^+=0$ with $p \cdot x = p^+ x^- - \un{p} \cdot \un{x}$. We have defined the following shorthand notation: 
\begin{equation}
\int\limits_{{p},\sigma}  \equiv \int \frac{\text{d}p^+ d^2 p_\perp}{(2\pi)^3 \, 2p^+} \sum_{\sigma=\pm} .
\end{equation}
The creation and annihilation operators are normalized such that
\begin{align}
[\hat{a}^a_{p,\lambda} , \hat{a}^{b \, \dagger}_{p',\lambda'}] = 2 p^+ \, (2 \pi)^3 \, \delta (p^+ - p'^+) \delta^2 (\un{p} - \un{p}') \, \delta^{ab} \, \delta_{\lambda \lambda'},
\end{align}
with the same expressions for the anti-commutators of the quark and anti-quark creation and annihilation operators.

The source current in the MV model is given by the large-$x$ partons. The source quarks lead to the current
\begin{align}\label{q_curr}
j^{a \, \mu}_{q} = g {\bar \psi}_0 \gamma^\mu \, t^a \, \psi_0 ,
\end{align}
with $t^a = t^a_F$ the fundamental generators of SU($N_c$).

To find the gluon source for the small-$x$ gluon field, let us consider the decomposition of the gauge field into
\begin{equation}\label{Aexp}
A^\mu = A_0^\mu + a^\mu,
\end{equation}
with $A_0^\mu$ the plane wave of the order-$g^0$ in the coupling due to the incoming or outgoing large-$x$ gluons in the nucleons, and $a^\mu$ the order-$g$ smaller-$x$ gluon field sourced by the large-$x$ gluons.
In Feynman gauge we have $\partial_\mu A^\mu = \partial_\mu A_0^\mu = \partial_\mu a^\mu = 0$. 
Starting from the Yang-Mills equations of motion (taken, for simplicity, without quark sources - those are considered separately at the sub-eikonal level since the end result for the gluon field is a linear superposition of the quark- and gluon-sourced terms)
\begin{align}
{\cal D}_\mu F^{\mu\nu}=0
\end{align}
and substituting \eq{Aexp} in it we obtain, at $\mathcal{O}(g^0)$,
\begin{align}\label{A0}
\Box A_0^{a, \mu} = 0
\end{align}
with $\Box$ the d'Alembert operator, while at $\mathcal{O}(g)$ we arrive at
\begin{equation}\label{Box_a}
\Box a^{a, \mu} = -gf^{abc} A_{0,\nu}^b F_0^{c,\nu\mu} - gf^{abc}\partial_\nu\left(A_0^{b,\nu}A_0^{c,\mu} \right),
\end{equation}
with $f^{abc}$ the SU($N_c$) structure constants. Equation \eqref{A0} is satisfied by the plane-wave field $A_0^\mu$ of the large-$x$ on-shell gluons. 
Acting with $\partial_\mu$ on the $\mathcal{O}(g)$ field-strength tensor for small-$x$ gluons, $f^{a,\mu\nu} = \partial^\mu a^{a,\nu} - \partial^\nu a^{a,\mu} + gf^{abc}A_0^{b,\mu} A_0^{c,\nu}$, and employing \eq{Box_a}, we obtain
\begin{align}\label{g_curr}
\partial_\mu f^{a,\mu\nu} = -gf^{abc}\left[ A_{0,\mu}^b\partial^\mu A^{c,\nu}_0 - A_{0,\mu}^b\partial^\nu A_0^{c,\mu} \right] = j_g^{a, \nu} .
\end{align}
As we will see below, on the right-hand side of \eq{g_curr} one can identify the first term as the current sourcing the sub-eikonal helicity-dependent field $\beta = f^{12}$ (for $\nu = \perp$, to be analyzed later in more detail) and the second term as the usual current sourcing the eikonal helicity-independent field $\alpha = a^+$ (for $\nu = +$). Both currents generating the gluon fields $\alpha$ and $\beta$ consist of the large-$x$ gluons inside the target nucleons.

In the eikonal approximation, for an $x^+$-direction moving nucleus, only $\mu = +$ component contributes in both the quark and gluon currents. The 3D color charge density operator is 
\begin{align}
\hat{\rho}^a_{cov} (x) = j^{a \, +} (x) \equiv j^{a \, +}_q (x) + j^{a \, +}_g (x).
\end{align}

Next we substitute the current \eqref{q_curr} into \eq{Odef2} and employ Eqs.~\eqref{psibd} and \eqref{psi_bar_bd}. We do not average over the quark colors and polarizations. In the eikonal approximation we also assume that $P^+_{ij} \gg \Delta^+_{ij}$. Suppressing the quark helicities, we obtain (keeping contributions corresponding to connected diagrams only)
\begin{align}
& \int \frac{d^2 \Delta_{ij}  d \Delta_{ij}^+ }{(2 \pi)^3 \, 2 P_{ij}^+} \, e^{-i \Delta_{ij}^+ \, b_{ij}^- + i \un{\Delta}_{ij} \cdot
    \un{b}_{ij}}   \notag \\  & \times \left\langle P_{ij}^+ + \thalf \Delta_{ij}^{+} , \thalf \un{\Delta}_{ij} \right| j^{a \, +}_{q} (x)  \left| -\thalf \un{\Delta}_{ij}, P_{ij}^+ - \thalf \Delta_{ij}^+ \right\rangle \notag \\  & =
\pm g \, t^a \, \delta (x^- - b_{ij}^-) \, \delta^2 (\un{x} - \un{b}_{ij})
\end{align}  
for the quark and anti-quark states (the former contribute with the plus sign, while the latter with the minus sign). This is in complete agreement with \eq{color_dens} if we denote $t^a_{\bar F} = - t^{a \, T}_F$.

Similar procedure for the gluon current \eqref{g_curr} yields, now with the gluon Fock states,
\begin{align}
& \int \frac{d^2 \Delta_{ij}  d \Delta_{ij}^+ }{(2 \pi)^3 \, 2 P_{ij}^+} \, e^{-i \Delta_{ij}^+ \, b_{ij}^- + i \un{\Delta}_{ij} \cdot
    \un{b}_{ij}}   \notag \\  & \times \left\langle P_{ij}^+ + \thalf \Delta_{ij}^{+} , \thalf \un{\Delta}_{ij} \right| j^{a \, +}_{g} (x)  \left| -\thalf \un{\Delta}_{ij}, P_{ij}^+ - \thalf \Delta_{ij}^+ \right\rangle \notag \\  & =
    g \, T^a \, \delta (x^- - b_{ij}^-) \, \delta^2 (\un{x} - \un{b}_{ij})
\end{align}  
for each large-$x$ gluon state contributing to $\rho^a$, with $T^a$ the adjoint generators of SU($N_c$), again in complete agreement with \eq{color_dens}. We see that the currents \eqref{q_curr} and \eqref{g_curr} indeed lead to \eq{color_dens} for the 3D color charge density. 

The above calculations can be streamlined if we work entirely in the coordinate space. To that end, define 
\begin{align}\label{x2p}
|\un{x}, x^- \rangle \equiv {\hat b}^\dagger_x |0\rangle
\end{align}
with 
\begin{align}\label{bdef}
{\hat b}^\dagger_x = \int \frac{d^2 p \, dp^+}{(2\pi)^3 \, \sqrt{2 p^+}} \, e^{- i p^+ x^- + i \un{p} \cdot \un{x}} \, {\hat b}^\dagger_p ,
\end{align}
such that
\begin{align}
\langle x^-, \un{x} | \un{y}, y^- \rangle = \delta (x^- - y^-) \, \delta^2 (\un{x} - \un{y}).
\end{align}
Note that the position eigenstate $|\un{x}, x^- \rangle$ is not boost-invariant unlike, say, the momentum eigenstate $| \un{p}, p^+ \rangle$. However, the state definition \eqref{x2p} is convenient for our purposes.

With the help of \eq{x2p} we rewrite \eq{Odef2} as
\begin{align}\label{Odef3}
& {\cal O} \left( b_{11}, \ldots , b_{A,N_A} \right) = \int \prod_{i=1}^{A} \prod_{j=1}^{N_i}  d^2 r_{ij} \, d r^-_{ij} \, e^{i P_{ij}^+ \, r^-_{ij}} \, \notag \\  & \times \prod_{i'=1}^A  \prod_{j'=1}^{{N_{i'}}} \left\langle b_{i'j'}^- + \thalf r_{i'j'}^{-} , \un{b}_{i'j'} +  \thalf \un{r}_{i'j'} \right| \hat {\cal O} \notag \\  & \times\prod_{i''=1}^A  \prod_{j''=1}^{{N_{i''}}} \left| \un{b}_{i''j''} -\thalf \un{r}_{i''j''}, b_{i''j''}^- - \thalf r_{i''j''}^- \right\rangle ,
\end{align}
now completely in coordinate space. In arriving at \eq{Odef3} we have assumed that $P^+_{ij} \gg \Delta^+_{ij}$. This is allowed since the typical $\Delta^+_{ij} \approx 1/b^-_{ij}$ with $b^-_{ij}$ spanning the full extend of the nucleus: hence  $P^+_{ij} \gg \Delta^+_{ij}$ for $A \gg 1$ and despite the Lorentz-contraction of the nucleus in the light cone minus direction.

Assuming that the momenta of large-$x$ partons' operators \eqref{ppA} contributing to the source currents all have approximately the same large ``plus" component $p^+$, we can re-write the eikonal contributions to the currents \eqref{q_curr} and \eqref{g_curr} as 
\begin{subequations}
\label{EQs:eik_Currents}
\begin{align}
& j^{a, \, \mu}_{q,\eik}(x^-,\un{x}) = \delta^\mu_+ \, g \, \sum_{\sigma=\pm}  \left[ \hat{b}^{i \, \dagger}_{x,\sigma} (t^a)_{ij} \hat{b}^j_{x,\sigma} - \hat{d}^{i \, \dagger}_{x,\sigma} (t^a)_{ji}  \hat{d}^j_{x,\sigma} \right], \\
& j^{a, \, \mu}_{g,\eik}(x^-,\un{x}) = \delta^\mu_+ \, g \, \sum_{\lambda=\pm} \hat{a}^{b , \, \dagger}_{x,\lambda} \, (T^a)_{bc} \, \hat{a}^c_{x,\lambda} ,
\end{align}
\end{subequations}
where ${\hat d}_{x, \sigma}$ and ${\hat a}_{x, \lambda}$ are defined by analogy to \eq{bdef}. Both currents are at $x^+ =0$. In writing the currents \eqref{EQs:eik_Currents} we have neglected the contributions which would lead to disconnected diagrams  when the currents are used in \eq{Odef3}. 

Let us define a combined eikonal source current by
\begin{equation}\label{jaa}
j_{\eik}^{a \, \mu} (x^-,\un{x}) = \delta^\mu_+ \, g \!\! \sum_{\substack{\sigma = \pm \\ r=\{q,\bar{q},g\}}}
\hat{\ag}_{r,x,\sigma}^\dagger \cdot t^a_r \cdot \hat{\ag}_{r,x,\sigma} ,
\end{equation}
where
\begin{equation}
\label{EQ:ac_op}
\hat{\ag}_{r, x, \sigma} = \begin{cases}
\hat{b}_{x, \sigma} & \text{for } r=q \\
\hat{d}_{x, \sigma} & \text{for } r=\bar{q} \\
\hat{a}_{x, \sigma} & \text{for } r=g
\end{cases}
\end{equation}
satisfy the equal light-cone-time commutation relation (see \eq{bdef})
\begin{equation}
\label{EQ:CR}
\left[ \hat{\ag}_{r,y,\sigma'}, \hat{\ag}^\dagger_{r,x,\sigma}\right]_r = \delta_{\sigma\sigma'}\ \delta (x^- - y^-) \, \delta^2 (\un{x} - \un{y}),
\end{equation}
with $[\ , \ ]_r$ being the commutator or anti-commutator depending on the irrep $r$. 

Substituting the current from \eq{jaa} as the operator into \eq{Odef3} we obtain
\begin{align}\label{Cave}
& \rho^a_{cov}(x^-, \un{x}) =  \int \prod_{i=1}^{A} \prod_{j=1}^{N_i}  d^2 r_{ij} \, d r^-_{ij} \, e^{i P_{ij}^+ \, r^-_{ij}} \, \notag \\  & \times \prod_{i'=1}^A  \prod_{j'=1}^{{N_{i'}}} \left\langle b_{i'j'}^- + \thalf r_{i'j'}^{-} , \un{b}_{i'j'} +  \thalf \un{r}_{i'j'} \right| j_{\eik}^{a \, \mu} (x^-,\un{x}) \notag \\  & \times\prod_{i''=1}^A  \prod_{j''=1}^{{N_{i''}}} \left| \un{b}_{i''j''} -\thalf \un{r}_{i''j''}, b_{i''j''}^- - \thalf r_{i''j''}^- \right\rangle 
\end{align}
which results in $\rho^a_{cov}$ given by \eq{color_dens}. As before, the matrix element in \eq{Cave} is non-diagonal in color space, so no color trace is implied. 

Anticipating the helicity-dependent case, it will be useful to define a fixed-helicity color current 
\begin{equation}\label{j_def_hel}
j^a_{\sigma}(x) =  g \sum_{r=\{q,\bar{q},g\}} 
\hat{\ag}_{r,x,\sigma}^\dagger \cdot t^a_r \cdot \hat{\ag}_{r,x,\sigma}
\end{equation}
such that the eikonal current is
\begin{equation}
j^{a \, \mu} _{\eik}(x) = \delta^\mu_+ \sum_{\sigma=\pm} j^a_{\sigma}(x) = \delta^\mu_+ \, \left[ j^a_{+}(x) + j^a_{-}(x) \right].
\end{equation}

An important ingredient of the MV model is the 2-dimensional color charge density \cite{McLerran:1993ni,McLerran:1993ka,McLerran:1994vd,Kovchegov:1996ty}
\begin{align}
\label{color_dens_2D}
\rho^a_{cov}(\un{x}) & \equiv \int\limits_{-\infty}^\infty d x^- \, \rho^a_{cov}(x^-, \un{x}) \\ & = g \sum_{i=1}^A \sum_{j=1}^{N_i} t_{R_{ij}}^a \, \delta^{2}(\un{x} -\un{b}_{ij}) . \notag
\end{align}


\subsection{Gaussian weight functional}
\label{ss:Gaussian_weight_functional}

The weight functional giving the distribution of the color charge density is Gaussian \cite{McLerran:1993ni,McLerran:1993ka,McLerran:1994vd}, as explicitly shown in \cite{Kovchegov:1996ty} for the density in \eq{color_dens}. Indeed, taking two color charge densities, $\hat{\rho}_{cov}^a (x) \, \hat{\rho}_{cov}^b (y) = j_{\eik}^{a \, +} (x) \, j_{\eik}^{b \, +} (y)$ with the currents from \eq{jaa}, inserting them into \eq{Odef3}, this time averaging over the helicities and colors, yields
\begin{align}\label{corr1}
 g^2 \sum_{i=1}^A \sum_{j=1}^{N_i} \delta^{(3)} {(\vecx - \vec{b}_{ij})} \delta^{(3)} (\vecy - \vec{b}_{ij}) \frac{ \text{tr} \left(  t_{R_{ij}}^a \,  t_{R_{ij}}^b \right)}{K_{R_{ij}}} \notag \\ = g^2 \, \delta^{(3)} (\vecx - \vecy) \sum_{i=1}^A \sum_{j=1}^{N_i} \delta^{(3)} {(\vecx - \vec{b}_{ij})}  \frac{C_{R_{ij}} \, \delta^{ab}}{K_A} , 
\end{align}
where we have defined $\vec{x} = (x^-,\xvec)$ and $\delta^{(3)} (\vecx) = \delta (x^-) \, \delta^2 (\un{x})$ for brevity. Here $K_{R_{ij}}$ is the dimension of the representation $R_{ij}$, such that $K_A = N_c^2 -1$ for the adjoint representation and $K_F = N_c$ for the fundamental one.  In \eq{corr1} we have simplified the color trace using
\begin{equation}
\frac{\text{tr} \left( t_R^a t_R^b \right)}{K_R} =  \delta^{ab} \frac{C_R}{K_A} \quad \text{for any irrep $R$},
\end{equation}
where $C_R$ is the Casimir operator of SU($N_c$) in the representation $R$. 

\begin{figure}[ht]
\begin{tikzpicture}[scale=.8]
\pgfmathsetseed{1} 
\foreach \i in {1,2.5,...,7} {
\draw[fill=black!40!white] (\i, rand) ellipse (.2 and 1.5);
};

\draw[fill=white,opacity=.7,white] (0,.2) rectangle (8,-.2);
\draw[thin] (0,.2) -- ++(8,0);
\draw[thin] (0,-.2) -- ++(8,0);
\draw[dashed] (0,0) -- ++(8,0);

\draw[<->] (8.2,.2) -- (8.2,-.2);
\node[right] at (8.2,0) {$r_\perp$};
\draw[<-] (7,.5) -- (8,1) node[right] {Nucleon};
\draw[<-] (7.5,-.05) to[out=-90,in=90] (9,-1) node[below] {Trajectory};

\draw[<-] (1,-.05) to[out=-120,in=180] (1,-2.7) node[right] {Volume probed};
\end{tikzpicture}
\caption{Sketch of a probe (dashed line) interacting with only a part of the large nucleus. The nucleons (darker gray) are randomly distributed in the nucleus. The light gray indicates the region that the probe ``sees" when crossing a given nucleon along its eikonal trajectory.}
\label{FIG:Cartoon}
\end{figure}
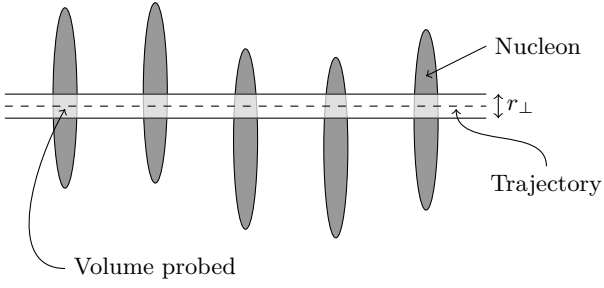

Note that the expression in \eq{corr1} is only non-zero when both $x$ and $y$ are equal: this way, both color density operators have to probe the same parton. In general this need not be the case: while certainly both $\hat{\rho}_{cov}^a (x)$ and $\hat{\rho}_{cov}^b (y)$ have to be in the same nucleon, they can still probe different partons in that nucleon. Such contributions give us a smooth function of $\un{x} - \un{y}$, not a $\delta$-function as in \eq{corr1}. Moreover, on the short transverse distance scales $r_\perp \ll 1/\Lambda_{QCD}$, which are of interest for perturbative QCD calculations at hand, the $\delta$-function is dominant due to our assumption that the large-$x$ source partons are spread out over non-perturbatively large distances from each other. In other words, as illustrated in \fig{FIG:Cartoon}, a given measurement in the perturbative region probes transverse distances $r_\perp \ll 1/\Lambda_{QCD}$, thus probing only a part of each nucleon. In this small fraction of the nucleon probed by a small probe, we are likely to find at most one parton. Therefore, the contribution captured in \eq{corr1} where both color charge densities connect to the same parton is dominant.

Substituting \eq{corr1} into \eq{eq:DMA5} we arrive at
\begin{align}
& \langle A | \rho_{cov}^a (x) \, \rho_{cov}^b (y) | A \rangle =  \delta^{(3)} (\vecx - \vecy) \sum_{i=1}^A \sum_{j=1}^{N_i} \frac{g^2 \, \delta^{ab}}{2 \, K_{R_{ij}}} \\ & \times \!\! \int \!\! d^2 B_i \, d B_i^- \frac{1}{A} \, \rho_A (B_i^-,
  \un{B}_i) \, \frac{1}{N_i} \, \rho_{N_i} (x^- - B_i^-, \un{x} -  \un{B}_i) . \notag
\end{align}
Next we assume that the parton number density in the nucleon $\rho_{N_i}$ is localized inside the nucleon and that the nucleon density $\rho_A$ is a slowly-varying function over this single-nucleon distance scale. In addition, for simplicity we assume that all the large-$x$ source partons are quarks, and that each nucleon contains the same number $N_q$ of them. 
We arrive at the following expression for the correlator of two color charge densities:
\begin{align}\label{corr2}
\langle A | \rho_{cov}^a (\un{x}, x^-) \, \rho_{cov}^b (\un{y}, y^-) | A \rangle & = \delta^{ab} \, \mu^2 (\un{x}, x^-)   \\ & \times \, \delta (x^- - y^-) \, \delta^2 (\un{x} - \un{y}), \notag
\end{align}
where
\begin{align}\label{mu3D}
\mu^2 (\un{x}, x^-) = 4 \pi \as \, \frac{N_q}{2 N_c} \, \rho_A (\un{x}, x^-) .
\end{align}
As usual, $\as = g^2/(4 \pi)$ is the QCD coupling constant. The generalization of this result to the case of a nucleus with a varying number of large-$x$ quarks and gluons in each nucleon is accomplished by the substitution
\begin{equation}
N_q \frac{1}{2N_c} \longrightarrow \left\langle N_R \, \frac{C_R}{K_A} \right\rangle
\end{equation}
where the angle brackets denote the averaging over all the nucleons in the nucleus and $N_R$ is the number of large-$x$ partons with color representation $R$  in each of the nucleons.

The correlator of two 2D color charge densities defined in \eq{color_dens_2D} is then
\begin{align}\label{corr3}
\langle A | \rho_{cov}^a (\un{x}) \, \rho_{cov}^b (\un{y}) | A \rangle & = \delta^{ab} \, \mu^2 (\un{x}) \, \delta^2 (\un{x} - \un{y})  
\end{align}
with
\begin{align}
\mu^2 (\un{x}) \equiv \int d x^- \, \mu^2 (\un{x}, x^-)  = 2 \pi \as \,  \frac{N_q}{N_c} \, T (\un{x}) 
\end{align}
and 
\begin{align}
T (\un{x}) \equiv \int\limits_{-\infty}^\infty d x^-  \, \rho_A (\un{x}, x^-)
\end{align}
the nuclear profile function. The quantity $\mu^2 (\un{x})$ has dimensions of momentum squared \cite{McLerran:1993ni,McLerran:1993ka,McLerran:1994vd,Kovchegov:1996ty} and is related to the quark saturation scale $Q_s$ in the MV model by
\begin{align}
Q_s^2 (\un{b}) = 2 \as C_F \, \mu^2 (\un{b}),
\end{align}
where $C_F = (N_c^2 -1)/(2 N_c)$ is the fundamental Casimir operator of SU($N_c$). Since $Q_s^2 \propto \mu^2 \propto A^{1/3}$ with $A$ the atomic number, the saturation scale is large for large nuclei, justifying the applicability of QCD perturbation theory.

Similar to the above calculation, one can show \cite{Kovchegov:1996ty} that the correlation function of any odd number of color charge densities \eqref{color_dens} is zero, while the correlation functions of an even number of densities reduce to a sum of terms factorized into products of two-density correlators \eqref{corr2}. For instance, for the four-density correlator one has
\begin{align}
& \langle A | \rho_{cov}^a (x_1) \, \rho_{cov}^b (x_2) \, \rho_{cov}^c (x_3) \, \rho_{cov}^d (x_4) | A \rangle  \\ & =  \langle A | \rho_{cov}^a (x_1) \, \rho_{cov}^b (x_2) | A \rangle \, \langle A | \rho_{cov}^c (x_3) \, \rho_{cov}^d (x_4) | A \rangle \notag \\ & + \mbox{permutations} . \notag
\end{align}
This implies that the functional distribution of the color charge density is Gaussian \cite{McLerran:1993ni,McLerran:1993ka,McLerran:1994vd,Kovchegov:1996ty}.  

Defining an abbreviated notation
\begin{align}
\langle \hcalo \rangle_A \equiv \langle A | \hcalo | A \rangle
\end{align} 
we note that in the MV model the expectation values of operators can be written as weighted functional averages
\begin{equation}
\langle \hcalo \rangle_A = \frac{\int \cald \rho_{cov} \ \hcalo[\rho_{cov}] \calw^{MV} [\rho_{cov}]}{\int \cald \rho_{cov} \ \calw^{MV} [\rho_{cov}]}
\end{equation}
with the weight functional $\calw^{MV}$. One usually normalized the weight functional to unity,
\begin{align}
\int \cald \rho_{cov} \ \calw^{MV} [\rho_{cov}] = 1,
\end{align}
such that 
\begin{equation}
\langle \hcalo \rangle_A = \int \cald \rho_{cov} \ \hcalo[\rho_{cov}] \calw^{MV} [\rho_{cov}] .
\end{equation}

Our above-mentioned results for the color charge density correlators imply that this weight functional is Gaussian and local \cite{McLerran:1993ni,McLerran:1993ka,McLerran:1994vd,Kovchegov:1996ty}, 
\begin{equation}\label{W_MV}
\mathcal{W}^{MV} [\rho_{cov}] \propto \text{exp} \left[ - \! \int \!  \text{d}^2 x_\perp d x^- \ \frac{\text{tr} \left[ \rho_{cov} (x) \cdot \rho_{cov} (x) \right]}{2\,\mu^2 (\un{x}, x^-) } \right].
\end{equation}
Here $\rho_{cov} = t^a \rho^a_{cov}$ with the trace in \eq{W_MV} going over the fundamental generators $t^a$.

One can use the classical equations of motion \eqref{Box_a} to write 
\begin{equation}
\Box a^+ = j^+_{\eik},
\end{equation}
to relate the eikonal current $j^+_{\eik}$ from \eq{jaa} and, hence, the 3D color charge density $\rho_{cov} = j^+_{\eik}$, to the eikonal gluon field $\alpha (\un{x}, x^-)  \equiv a^+ (\un{x}, x^-) $. Since the field does not depend on $x^+$ the d'Alembertian reduces to $-\nabla_\perp^2$, where $\nabla_\perp^2$ is the 2D Laplacian in the transverse plane. We obtain
\begin{equation}\label{Box_alpha}
\Box \alpha = - \nabla^2_\perp \alpha =  j_{\eik}^+ = \rho_{cov}.
\end{equation}
The MV weight functional can be re-written in terms of the eikonal field $\alpha$ as
\begin{align}\label{WMVa}
\mathcal{W}^{MV} [\alpha] \propto \text{exp} \left[ - \! \int \! \text{d}^2 x_\perp d x^- \ \frac{\text{tr} \left[ \nabla^2_\perp \alpha(x) \cdot \nabla^2_\perp \alpha(x) \right] }{2\,\mu^2 (\un{x}, x^-)} \right].
\end{align}
This concludes the derivation of the MV model target weight functional.


\section{Generalization to helicity-dependent case}
\label{sec:Gen_hel}

From the above discussion one expects that the weight functional remains Gaussian after the sub-eikonal fields are included, as required in generalizing the calculation to include helicity effects: indeed this Gaussianity is ensured by the large color-neutral nucleus in the MV model consisting of many independent color-neutral nucleons. The Gaussian distribution has to remain local, since the same parton has to be the source of two color charge densities in the two-point correlator even at the sub-eikonal order. For higher-order correlators in a large nucleus, the dominant contribution again comes from the pairs of color charge densities coming from partons in different nucleons (while each pair comes from the same parton). Below we see in detail how this Gaussian distribution emerges for the sub-eikonal fields $\beta$, $\psi$/$\bpsi$. 
We work in covariant gauge with the results also valid in $A^-=0$ gauge.


\begin{widetext}

\subsection{Helicity-dependent operators}

\begin{figure}[ht!]
\begin{center}
\includegraphics[width= 0.93 \textwidth]{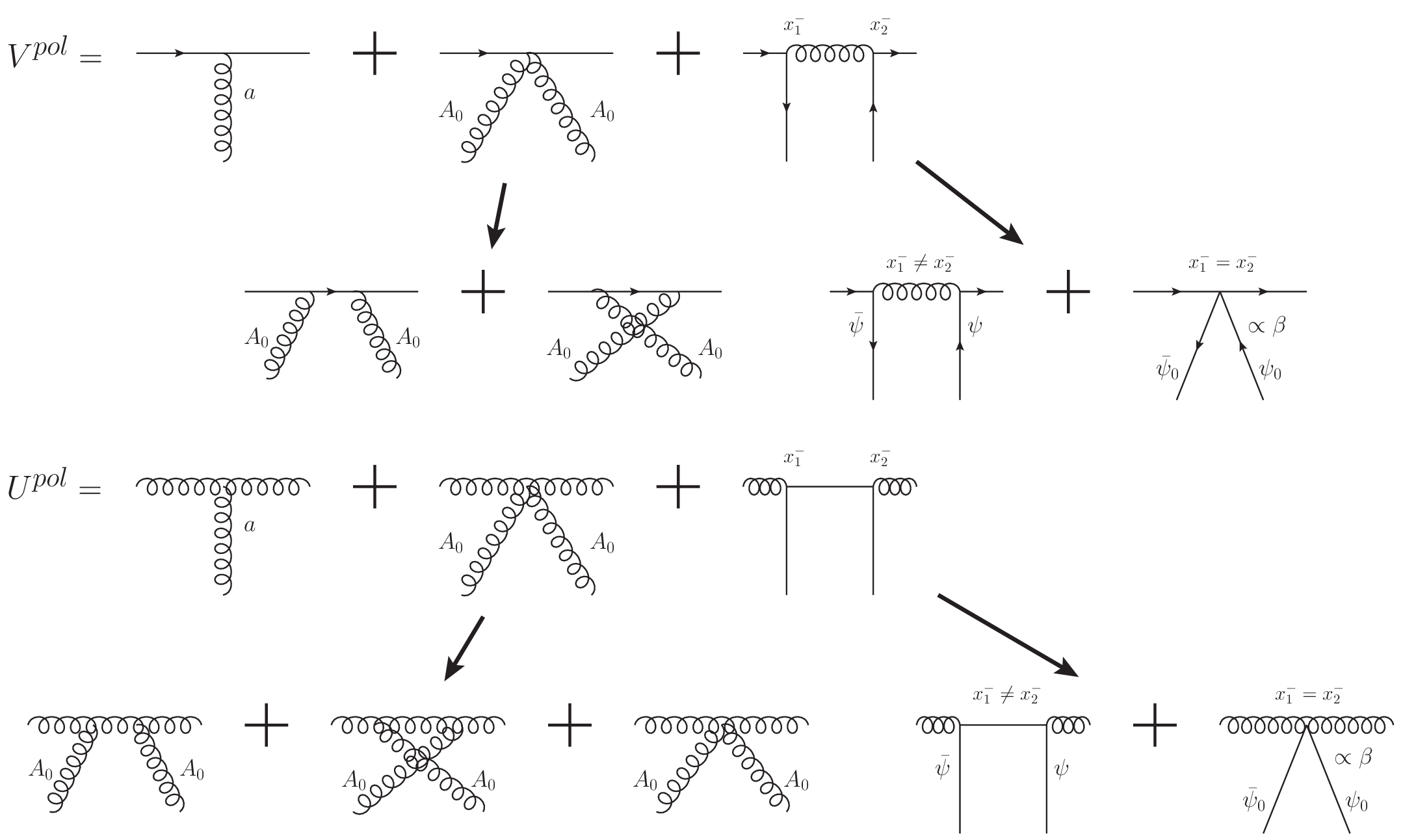} 
\caption{Sub-eikonal interactions included in the polarized Wilson lines with a generic target composed of both quarks and gluons at large $x$. Horizontal lines indicate propagation of the quark or gluon probe along the light-cone minus direction, while the non-horizontal lines indicate the background fields of the target. The sub-eikonal field $\beta$ is defined in \eq{beta_def}, and receives contributions from the plane-wave field $A_0^\mu$ of the large-$x$ gluons and the field $a^\mu$ of the small-$x$ gluons, as per the decomposition in \eq{Aexp}. The diagrammatic structure of the $A_0 \, A_0$ contribution is indicated by the left set of arrows (see Appendix~\ref{App:A} for details). The right set of arrows shows how the interaction with the quark fields consists of the non-local contribution ($x_1^- \neq x_2^-$) due to the small-$x$ quark fields $\psi$ and $\bpsi$, and the local term $\sim \bpsi_0 (x) \gamma^+ \gamma^5 \psi_0 (x)$ due to the large-$x$ quark fields $\psi_0 / \bpsi_0$, giving a contribution proportional to $\beta$, as elucidated in Appendix~\ref{App:B}.}
\label{FIG:Subeik_int_vertices}
\end{center}
\end{figure}

Helicity-dependent interactions of a probe with the background field generated by the polarized target are given by the sub-eikonal vertices were calculated in \cite{Kovchegov:2017lsr,Kovchegov:2018znm} resulting in the so-called quark and gluon ``polarized Wilson lines"
\begin{align} 
  \label{eq:Wpol_all} 
  & V^{pol}_{\un x} = \frac{i g p_1^+}{s} \,
  \int\limits_{-\infty}^\infty d x^- \, V_{\ul x} [+\infty, x^-] \: 
  F^{12} (x^-, \un{x}) \: V_{\ul x} [x^- , -\infty] \\ & - \frac{g^2 \, p_1^+}{s}
  \int\limits_{-\infty}^\infty d x_1^- \, \int\limits_{x_1^-}^\infty d x_2^- \, V_{\ul x} [+\infty, x_2^-] \:  t^b \, {\psi}_\beta (x_2^-, {\un x})  \, U_{\ul x}^{ba} [ x_2^-,  x_1^-] \left[ \frac{1}{2} \, \gamma^+ \, \gamma^5 \right]_{\alpha\beta} \, {\bar \psi}_\alpha (x_1^-, {\un x}) \, t^a  \: V_{\ul x} [x_1^- , -\infty] \notag 
\end{align}
and
\begin{align} 
  \label{M:UpolFull}
  & (U_{\ul x}^{pol})^{ab} = \frac{2 i \, g \, p_1^+}{s}
  \int\limits_{-\infty}^{+\infty} dx^- \: \left( U_{\ul{x}}[+\infty, x^-] \:
  {\cal F}^{12} (x^+ =0 , x^- , \ul{x}) \: U_{\ul{x}} [x^- , -\infty] \right)^{ab} \\ &  - \frac{g^2 \, p_1^+}{s} \, \int\limits_{-\infty}^\infty d x_1^- \, \int\limits_{x_1^-}^\infty d x_2^- \, U^{aa'}_{\un x} [+\infty, x_2^-] \,  {\bar \psi} (x_2^-, {\un x}) \, t^{a'} \, V_{\un x} [x_2^-, x_1^-] \, \frac{1}{2} \, \gamma^+ \gamma_5 \, t^{b'} \,  \psi (x_1^-, {\un x}) \, U^{b'b}_{\un x} [x_1^-, -\infty] - c.c.   \notag
\end{align}
for the quark and gluon interaction with the polarized nucleus. 
\end{widetext}
Here, $p_1^+$ is the large light-cone momentum of the interacting parton in the nucleus,  $s$ is the center-of-mass scattering energy squared between the quark or gluon in the projectile and the above-mentioned parton, while the light-cone Wilson lines are denoted by
\begin{align}
  V_{\un{x}} [b^-, a^-] = \mathcal{P} \exp \left[ i g
    \int\limits_{a^-}^{b^-} d x^- \, A^+ (x^+=0, x^-, {\un x})
  \right],
\end{align}
for the fundamental representation and by
\begin{align}
  U_{\un{x}} [b^-, a^-] = \mathcal{P} \exp \left[ i g
    \int\limits_{a^-}^{b^-} d x^- \, {\cal A}^+ (x^+=0, x^-, {\un x})
  \right]
\end{align}
for the adjoint representation. (Calligraphic notation for the field ${\cal A}^+$ and field strength ${\cal F}^{12}$ indicate adjoint representation.)

While the ``polarized Wilson lines" \eqref{eq:Wpol_all} and \eqref{M:UpolFull} were derived in \cite{Kovchegov:2017lsr,Kovchegov:2018znm} by assuming that all interactions are separated in $x^-$, the resulting operators \eqref{eq:Wpol_all} and \eqref{M:UpolFull} apply beyond this assumption. The sub-eikonal interactions of the quark and gluon probe with the longitudinally polarized target containing both quarks and gluons at large-$x$ are shown in \fig{FIG:Subeik_int_vertices}. As can be seen from Eqs.~\eqref{eq:Wpol_all} and \eqref{M:UpolFull}, in addition to the gluon eikonal field $\alpha= A^+=a^+$, we also have to consider the sub-eikonal helicity-depend gluon field in Feynman gauge \cite{Kovchegov:2017lsr,Kovchegov:2018znm,Cougoulic:2019aja} 
\begin{equation}\label{beta_def}
\beta^a = F^a_{12}.
\end{equation} 
Helicity information can also be exchanged between the probe and the target by quark exchanges, thus we need the quark and anti-quark fields $\psi$ and $\bar{\psi}$, in addition to the two gluon fields $\alpha$ and $\beta$, as follows from Eqs.~\eqref{eq:Wpol_all} and \eqref{M:UpolFull} as well.
Those fields are used to build the helicity-dependent small-$x$ evolution equation in \cite{Cougoulic:2019aja,Kovchegov:2018znm}.

As one can see from \fig{FIG:Subeik_int_vertices}, in the sub-eikonal polarization-dependent case at hand, the interaction between the projectile and target is not limited to $t$-channel exchanges, and includes $s$- and $u$-channel processes as well. In \fig{FIG:Subeik_int_vertices} we briefly sketch how those $s$- and $u$-channel processes, along with the 4-gluon vertex interaction, enter the expressions \eqref{eq:Wpol_all} and \eqref{M:UpolFull}, with more details given in Appendices~\ref{App:A} and \ref{App:B}. The presence of the $s$- and $u$-channel processes makes the standard MV-model separation of the entire interaction into a projectile scattering on the small-$x$ field of the target somewhat difficult. We see that in the $s$- and $u$-channel interactions with the gluon target the projectile couples to the large-$x$ target field $A^0$ directly (see Appendix~\ref{App:A}). Similar direct coupling to the large-$x$ quark fields $\psi_0 / \bpsi_0$, defined in more detail below in \eq{psi0}, is possible for the quark target (see Appendix~\ref{App:B}). At the same time, the target density should be the starting point of our calculation, since, as we mentioned above, the Gaussian nature of the weight functional (as in \eq{W_MV}) persists for the sources beyond the eikonal approximation, and is simply a consequence of the target being a large nucleus made out of color-neutral nucleons.

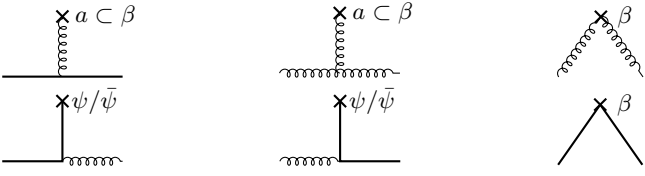
\begin{figure}[ht!]
\begin{tikzpicture}[scale=.8]
\draw[thick] (0,0) -- ++ (2,0);
\draw[gluon] (1,0) -- ++(0,1) node[right] {$\, a \subset \beta$};
\draw[thick] (.9,.9) -- ++(.2,.2);
\draw[thick] (.9,1.1) -- ++(.2,-.2);
\end{tikzpicture}
\hfill
\begin{tikzpicture}[scale=.8]
\draw[gluon] (0,0) -- ++ (2,0);
\draw[gluon] (1,0) -- ++(0,1) node[right] {$\, a \subset \beta$};
\draw[thick] (.9,.9) -- ++(.2,.2);
\draw[thick] (.9,1.1) -- ++(.2,-.2);
\end{tikzpicture}
\hfill
\begin{tikzpicture}[scale=.8]
\draw[gluon] (.3,0) to (1,1);
\draw[gluon] (1,1) to (1.7,0);
\node[right] at (1,1) {$\ \beta$};
\draw[thick] (.9,.9) -- ++(.2,.2);
\draw[thick] (.9,1.1) -- ++(.2,-.2);
\end{tikzpicture}
\hfill
 
\begin{tikzpicture}[scale=.8]
\draw[thick] (0,0) -- ++ (1,0) -- ++(0,1) node[right] {$\psi/\bar{\psi}$};
\draw[gluon] (1,0) -- ++(1,0);
\draw[thick] (.9,.9) -- ++(.2,.2);
\draw[thick] (.9,1.1) -- ++(.2,-.2);
\end{tikzpicture}
\hfill
\begin{tikzpicture}[scale=.8]
\draw[thick] (2,0) -- ++ (-1,0) -- ++(0,1) node[right] {$\psi/\bar{\psi}$};
\draw[gluon] (0,0) -- ++(1,0);
\draw[thick] (.9,.9) -- ++(.2,.2);
\draw[thick] (.9,1.1) -- ++(.2,-.2);
\end{tikzpicture}
\hfill
 \begin{tikzpicture}[scale=.8]
\draw[thick] (.3,0) to (1,1);
\draw[thick] (1,1) to (1.7,0);
\node[right] at (1,1) {$\ \beta$};
\draw[thick] (.9,.9) -- ++(.2,.2);
\draw[thick] (.9,1.1) -- ++(.2,-.2);
\end{tikzpicture}
\hfill

\caption{The topologies of sub-eikonal vertices required to generate the sub-eikonal fields $\beta$, $\psi$, and $\bpsi$. Horizontal and diagonal lines indicate the propagation of a source in the target along the light-cone plus direction. Each vertical line indicates a propagator from the source (the vertex point) to the position where the field is measured (denoted by the cross). Straight lines are either in the fundamental irrep or in its conjugate.}
\label{FIG:Subeik_sources}
\end{figure}

With this in mind, we observe that the interactions in \fig{FIG:Subeik_int_vertices} can be separated into the projectile coupling to $\beta, \psi$ and $\bpsi$. Below, we focus on generating those fields from the target. Above, in \eq{g_curr} we saw that the gluon field-strength operator $\beta$ can be written as sourced by a particular gluon (or quark) current. The interaction vertices of interest yielding the fields $\beta, \psi$ and $\bpsi$ are given in \fig{FIG:Subeik_sources}. The approach we employ below is similar to that for the eikonal gluon field reviewed in the previous Section. We will start from the source operator, and work our way to the weight functional for the sub-eikonal fields.


\subsection{Source operators}

\paragraph{Sub-eikonal source: gluon field strength.}

In order to generate the sub-eikonal gluon field strength $\beta$, we need a source current as depicted by the three diagrams in the top row of \fig{FIG:Subeik_sources} along with the right-most diagram in the bottom row of the same figure. Let us detail only the case of a quark source current since the cases of the anti-quark and gluon sources would follow almost automatically from it. This leaves us with the left-most diagram in the top row of \fig{FIG:Subeik_sources} and with the right-most diagram in the bottom row. Here we will concentrate only on the former: the evaluation of the latter is carried out in Appendix~\ref{App:B}.

We are interested in the operator $g \bpsi_0 \gamma^\mu t^a \psi_0$ (see \eq{q_curr}). The eikonal contribution was simply obtained for $\mu=+$, we are now interested in the sub-eikonal helicity-dependent contribution given by $\mu = \perp$. This component gives the leading helicity-dependent contribution to  $g \bpsi_0 \gamma^\mu t^a \psi_0$ from \eq{q_curr}. Suppressing the anti-quark contribution for now we get
\begin{align}
&j^{a \, \mu}_{quarks,sub\eik} = \!\!\!\!\!
\int\limits_{{q,\sigma'; p,\sigma}} \!\!\!\! g \, \hat{b}^{i \, \dagger}_{q,\sigma'} (t^a)_{ij} \hat{b}^j_{p,\sigma} \ \bar{u}_{\sigma'}(q) \gamma^\mu_\perp u_\sigma(p) e^{ix\cdot(q-p)} \notag \\
&\simeq \! - g \!\! \sum_{\sigma=\pm} \! \sigma \epsilon_\perp^{\mu i}\partial^i_\perp \!\! \int \! \frac{d^2 q_\perp d q^+ \, d^2 p_\perp d p^+}{(2\pi)^6 2 q^+ \, 2 p^+}  \hat{b}^{i \, \dagger}_{q,\sigma} (t^a)_{ij} \hat{b}^j_{p,\sigma} \, e^{ix\cdot(q-p)} \notag \\
&\simeq - g \sum_{\sigma=\pm}\sigma \frac{1}{2\langle p^+ \rangle} \,\epsilon_\perp^{\mu i}\partial^i_\perp \ \hat{b} ^{i \, \dagger}_{x,\sigma} (t^a)_{ij}  \hat{b}_{x,\sigma}^j  \label{jq1}
\end{align}
with $\epsilon_\perp^{\mu i}$ the two-dimensional Levi-Civita symbol. Starting from the second line of \eq{jq1} we have only kept the leading helicity-dependent contribution (i.e., $\sigma \, \delta_{\sigma\sigma'}$) and replaced the transverse momenta by a derivative using $i(q-p)^i e^{ix\cdot(q-p)} = \partial_\perp^i e^{ix\cdot(q-p)}$. We have also used Brodsky-Lepage spinors \cite{Lepage:1980fj} (see Eqs.~\eqref{BLspinors} below). In the last line of \eq{jq1} we have anticipated the fact that the sources have a large $p^+$, and factored out $1/p^+$ as a smooth function compared to the rapidly oscillating $e^{ip^+x^-}$. Furthermore we have assumed that all sources have approximately the same ``plus" momentum $\langle p^+ \rangle$ which led to the sub-eikonal suppression $\sim \langle p^+ \rangle^{-1}$. This assumption simplifies the algebra, but is not necessary: $\langle p^+ \rangle$ can be replaced by an operator assuming different (large) $p^+$ values when acting on different large-$x$ parton states in different nucleons.

Introducing back the contributions from the anti-quarks and gluons (using \eq{g_curr}) in the target while still working in covariant gauge yields

\begin{align}
\label{EQ:Source_subeik_gluon}
j^{a \, \mu}_{subeik}(x^-, {\un x}) &\simeq 
\ - g \frac{\epsilon_\perp^{\mu i}\partial^i_\perp}{2 \langle p^+ \rangle} 
\sum_{\substack{\sigma = \pm \\ r=\{q,\bar{q},g\}}}
\sigma  \,  \hat{\ag}^\dagger_{r, x,\sigma} \cdot t^a_r \cdot \hat{\ag}_{r, x,\sigma} .
\end{align}
This expression agrees with the diagrammatic analysis carried out in Appendices~\ref{App:A} and \ref{App:B} and provides the source current corresponding to all diagrams in the top line of \fig{FIG:Subeik_sources}. There is one important caveat: in the case of a flavor-singlet scattering, which is what we consider here, the right-most diagram in the bottom row of \fig{FIG:Subeik_sources} gives a non-zero contribution, though only for the case when the probe is a gluon. The contribution of this diagram (with the gluon probe) is accounted for by reducing the result in \eq{EQ:Source_subeik_gluon} by 2. We will keep this in mind for the later calculations. 

Employing the definite-helicity color current \eqref{j_def_hel}
we rewrite the current in \eq{EQ:Source_subeik_gluon} as
\begin{equation}\label{j_subeik_gluon}
j^{a \, \mu}_{subeik} (x^-, {\un x}) = - \frac{\epsilon_\perp^{\mu i}\partial^i_\perp}{2\langle p^+ \rangle} \sum_{\sigma = \pm} \sigma \, j^a_\sigma(x^-, {\un x}).
\end{equation}
This expression will be useful latter when we construct the target weight functional involving the $\beta$-field.


\paragraph{Sub-eikonal source: quark fields.}

To construct a source for the quark and anti-quark fields, it is useful to first define a way to decompose the 4-dimensional space of spinors. The Brodsky-Lepage (BL) spinors \cite{Lepage:1980fj} are defined by 
\begin{subequations}\label{BLspinors}
\begin{align}
& u_\sigma (p) = \frac{1}{\sqrt{\sqrt{2} \, p^+}} \, [\sqrt{2} \, p^+ + m \, \gamma^0 +  \gamma^0 \, {\un \gamma} \cdot {\un p} ] \,  \chi_\sigma, \\ & v_\sigma (p) = \frac{1}{\sqrt{\sqrt{2} \, p^+}} \, [\sqrt{2} \, p^+ - m \, \gamma^0 +  \gamma^0 \, {\un \gamma} \cdot {\un p} ] \,  \chi_{-\sigma},
\end{align}
\end{subequations}
with
\begin{equation}
\chi_+ = \frac{1}{\sqrt{2}}
\begin{pmatrix}
1\\ 0\\ 1\\ 0
\end{pmatrix},
\qquad 
\chi_- = \frac{1}{\sqrt{2}}
\begin{pmatrix}
0\\ 1\\ 0\\ -1
\end{pmatrix}.
\end{equation}
Those spinors are useful to describe plus-direction moving particles, in our case, the partons inside the target. In addition to the BL spinors, it is convenient to define the ``anti-BL" spinors \cite{Kovchegov:2018znm,Kovchegov:2018zeq}, obtained from the BL spinors by interchanging the plus and minus directions, as
\begin{subequations}
\begin{align}\label{Eq:antispinnors_def_a}
& u^{anti}_\sigma (p) = \frac{1}{\sqrt{\sqrt{2} \, p^-}} \, [\sqrt{2} \, p^- + m \, \gamma^0 +  \gamma^0 \, {\un \gamma} \cdot {\un p} ] \,  \rho_\sigma, \\ 
\label{Eq:antispinnors_def_b}
& v^{anti}_\sigma (p) = \frac{1}{\sqrt{\sqrt{2} \, p^-}} \, [\sqrt{2} \, p^- - m \, \gamma^0 +  \gamma^0 \, {\un \gamma} \cdot {\un p} ] \,  \rho_{-\sigma},
\end{align}
\end{subequations}
with
\begin{equation}
\rho_+ = \frac{1}{\sqrt{2}}
\begin{pmatrix}
1\\ 0\\ -1\\ 0
\end{pmatrix},
\qquad 
\rho_- = \frac{1}{\sqrt{2}}
\begin{pmatrix}
0\\ 1\\ 0\\ 1
\end{pmatrix}.
\end{equation}
Those spinors are useful to describe minus-moving particles, that is, particles moving in the direction of the probe in our case.
The spinors $\chi_\sigma$ and $\rho_\sigma$ are simply related to each other with the help of the Dirac matrix $\gamma^0$ (in the Dirac representation):
\begin{equation}\label{gamma0}
\rho_\pm=\gamma^0\chi_\pm = \gamma^0 (\gamma^0\rho_\pm).
\end{equation}

We know from the diagrammatic analysis of \cite{Kovchegov:2018znm} that the dominant high-energy (helicity-dependent and helicity-independent) interaction via a $t$-channel quark exchange between the probe and the target (see \fig{FIG:Subeik_int_vertices} and the left two diagrams in the bottom row of  \fig{FIG:Subeik_sources}) involves  $\rho_{\sigma_p}^T \gamma^0 \psi$ and $\bpsi \rho_{\sigma_p} = (\rho_{\sigma_p}^T \gamma^0 \psi)^\dagger$, where $\sigma_p$ is the helicity of the incoming or outgoing quark probe. 
Using \eq{gamma0} we write
\begin{align}\label{proj_chi}
\rho_{\sigma_p}^T \gamma^0 \psi 
=\chi_{\sigma_p}^T \slashed{\Delta} j_\psi ,
\end{align}
where $\slashed{\Delta} \equiv [i \slashed{\pd} - m]^{-1}$ is the inverse Dirac operator for the quark field, such that the solution of the linearized Dirac equation 
\begin{align}\label{Eq:Psi_EoM}
[i \slashed{\pd} - m] \psi = j_\psi  
\end{align} 
with the (sub-eikonal) source current
\begin{align}\label{Jpsi}
j_\psi = - g \, \slashed{A} \, \psi
\end{align}
can be written as
\begin{align}\label{psi0}
\psi = \psi_0 + \slashed{\Delta} j_\psi .
\end{align}
Here $\psi_0$ is the field of large-$x$ quarks in the target, which is the quark analogue of the gluon field $A_0^\mu$. It solves the free Dirac equation, $[i \slashed{\pd} - m] \psi_0 =0$, and contributes to both Figures~\ref{FIG:Subeik_int_vertices} and \ref{FIG:Subeik_sources} via $s$- and $u$-channel exchanges, as clarified in Appendix~\ref{App:B}. By contrast, the $\slashed{\Delta} j_\psi$ term gives the small-$x$ quark field, contributing to the $t$-channel quark exchanges.

The sub-eikonal source current \eqref{Jpsi} for the quark field receives contributions from the quark and gluon lines at the bottom of the left two panels in the lower row of Fig.~\ref{FIG:Subeik_sources}. Using the field operator definition \eqref{Jpsi} and expanding it to order-$g$ by using $A_0$ and $\psi_0$ in it one arrives at
\begin{align}\label{EQ:JPsi_proj}
\!\! j_\psi^i (x^-, \un{x}) & = -g \!\!\! \int\limits_{{p,\lambda;q,\sigma}}  \!\!\! \bigg\{
\hat{a}^{a \, \dagger}_{p,\lambda} (t^a)_{ik} \hat{b}^k_{q,\sigma} \, e^{ix\cdot(p-q)}
\left[ \slashed{\epsilon}_\lambda^* (p) u_\sigma(q)\right] \nonumber \\
+ & (t^a)_{ik}
\hat{d}^{k \, \dagger}_{q,\sigma} \, \hat{a}^a_{p,\lambda}  \, e^{-ix\cdot(p-q)}
\left[ \slashed{\epsilon}_\lambda (p) v_\sigma(q)\right] \bigg\} , 
\end{align}
where we only keep the terms with the incoming quark multiplying the outgoing gluon and the incoming gluon multiplying the outgoing anti-quark. These are the terms shown in the left two diagrams in the bottom row of \fig{FIG:Subeik_sources}. The terms we neglect in arriving at \eq{EQ:JPsi_proj} are suppressed in the averaging \eqref{Odef2} for the high-$p^+$ and $q^+$ sources, if we are interested in the small-$x$ quark field. Note that the source current $j_\psi^i$ comes with the fundamental color index $i$. In addition, $j_\psi^i$ is a Dirac spinor. 

Due to \eq{proj_chi}, we are interested only in the $\chi_{\sigma_p}^T \slashed{\Delta} j_\psi^i$ projection of the current $j_\psi^i$ from \eq{EQ:JPsi_proj}, since this is the part of the current relevant for high energy scattering, including helicity transfer. This is perhaps analogous to the fact that only a particular projection of the gluon field given by $\beta$ in \eq{beta_def} is relevant for the sub-eikonal helicity-dependent gluon exchanges. 

One can define four projectors to span the full spinor space,
\begin{equation}
\mathcal{P}_{\chi,\pm} \equiv \chi_\pm \chi_\pm^T, \qquad \mathcal{P}_{\rho,\pm} \equiv \rho_\pm \rho_\pm^T.
\end{equation}
Those projectors have the property of idempotency, orthogonality, and completeness. Employing the completeness relation for the projectors, one writes
\begin{align}
\chi_{\sigma_p}^T \slashed{\Delta} j_\psi^i = \sum_{k \in\{\chi,\rho\}\times\{+,-\}} \chi_{\sigma_p}^T \, \slashed{\Delta} \,\mathcal{P}_{k} \, j_\psi^i . 
\end{align}
For simplicity consider the massless case, $m=0$, where $\slashed{\Delta} = - i \slashed{\pd} /\pd^2$.
Since $\gamma^- \chi_{\sigma} =0$, $\gamma^+ \rho_{\sigma} =0$, $\chi_{\sigma}^T \gamma^+ =0$ and $\rho_{\sigma}^T \gamma^- =0$ we see that
\begin{align}
\chi_{\sigma_p}^T \, \slashed{\Delta} \, j_\psi^i & = - i \chi_{\sigma_p}^T \, \frac{ \pd^+ \gamma^-  - {\un \pd} \cdot {\un \gamma} }{\pd^2} \, j_\psi^i  \notag \\ & \approx - i \chi_{\sigma_p}^T \, \frac{{\un \pd} \cdot {\un \gamma} }{\pd_\perp^2} \, j_\psi^i 
\end{align}
since $\pd^+ j_\psi^i \propto (p^+ - q^+)$, which is suppressed by a power of energy compared to the ${\un \pd} \cdot {\un \gamma}$ term. Reinstating the projectors we have
\begin{align}\label{curr_proj4}
\chi_{\sigma_p}^T \slashed{\Delta} j_\psi^i =  \sum_{k \in\{\chi,\rho\}\times\{+,-\}} (- i) \, \chi_{\sigma_p}^T \, \frac{{\un \pd} \cdot {\un \gamma} }{\pd_\perp^2} \, \,\mathcal{P}_{k} \, j_\psi^i . 
\end{align}
Further, observing that for $i=1,2$
\begin{subequations}
\begin{align}
& \chi_{\sigma_1}^T  \gamma^i \chi_{\sigma_2} = (- \sigma_1 \, \delta^{i1} + i \, \delta^{i2} ) \, \delta_{\sigma_1, - \sigma_2}, \\
& \chi_{\sigma_1}^T  \gamma^i \rho_{\sigma_2} = 0,
\end{align}
\end{subequations}
we conclude that only the $\mathcal{P}_{\chi,\pm}$ projectors contribute in the sum of \eq{curr_proj4}, such that
\begin{align}\label{curr_proj5}
\chi_{\sigma_p}^T \slashed{\Delta} j_\psi^i =  \sum_{\sigma} (- i) \, \chi_{\sigma_p}^T \, \frac{{\un \pd} \cdot {\un \gamma} }{\pd_\perp^2} \, \chi_\sigma \, \chi_\sigma^T  \, j_\psi^i . 
\end{align}
We see that only the $\chi_\sigma^T  \, j_\psi^i$ projection of the current \eqref{EQ:JPsi_proj} contributes to the high-energy (helicity-dependent and helicity-independent) interactions that we are interested in here. 

Denoting this projection 
\begin{align}\label{Jdef}
J_\psi^{i \, \sigma} \equiv \chi_\sigma^T  \, j_\psi^i
\end{align}
 and expanding it to order-$g$ by using $A_0$ and $\psi_0$ in it, we find, after some algebra, 
\begin{align}\label{curr_proj6}
J_\psi^{i \, \sigma} (x^-, \un{x}) & \simeq \frac{-g}{\sqrt{\sqrt{2} \, \langle p^+ \rangle}} \\ & \times \, \left[ 
\hat{a}^{a \, \dagger}_{x,-\sigma} \, (t^a)_{ik} \, \hat{b}^k_{x,-\sigma}   
- (t^a)_{ik} \, \hat{d}^{k \, \dagger}_{x, -\sigma} \, \hat{a}^a_{x,\sigma} \right] . \notag
\end{align}

The source current for the small-$x$ anti-quark field is 
\begin{align}
{\bar j}_{\psi} = - g {\bar \psi} \, \slashed{A} = j_\psi^\dagger \, \gamma^0. 
\end{align}
Its relevant projection $J_{\bar \psi}^{i \, \sigma}  \equiv  {\bar j}^i_{\psi} \, \rho_\sigma =  \left(  \chi_\sigma^T j^i_{\psi} \right)^\dagger$ at order-$g$ is given by the hermitean conjugate of the current projection in \eq{curr_proj6}, 
\begin{align}\label{curr_proj7}
J_{\bar \psi}^{i \, \sigma} & = \left( J_\psi^{i \, \sigma} \right)^\dagger \simeq \frac{-g}{\sqrt{\sqrt{2} \, \langle p^+ \rangle}} \\ & \times \, \left[ 
\hat{a}^a_{x,-\sigma} \, (t^a)_{ki} \, \hat{b}^{k  \, \dagger}_{x,-\sigma}  
- (t^a)_{ki} \, \hat{d}^k_{x, -\sigma} \, \hat{a}^{a  \, \dagger}_{x,\sigma} \right] . \notag
\end{align}


\subsection{Weight functional for the gluon fields}

Let us now construct the weight functional for the helicity-dependent MV model. We start in the gluon sector, working with the $\alpha(x)$ and $\beta (x)$ fields. Similar to the MV model derivation done in Sec.~\ref{ss:Gaussian_weight_functional}, one can start with the fixed-helicity currents in \eq{j_def_hel}  (instead of the eikonal currents) and show that their distribution is indeed Gaussian. The derivation is completely analogous to Sec.~\ref{ss:Gaussian_weight_functional} and will not be repeated here in detail. One can show that the two-point function is proportional to the three-dimensional delta-function,
\begin{align}\label{corr_sigma}
\langle A | j_{\sigma}^a (\un{x}, x^-) \, j_{\sigma'}^b (\un{y}, y^-) | A \rangle & = \delta^{ab} \, \delta_{\sigma \sigma'} \, \mu_\sigma^2 (\un{x}, x^-)   \\ & \times \, \delta (x^- - y^-) \, \delta^2 (\un{x} - \un{y}), \notag
\end{align}
with 
\begin{align}\label{mu3Dsigma}
\mu^2_\sigma (\un{x}, x^-) = 2 \pi \as \, \frac{N_q^\sigma}{N_c} \, \rho_A (\un{x}, x^-) .
\end{align}
Again, for simplicity we have assumed that the large-$x$ sources in the nucleons are (``valence") quarks, where $N_q^\sigma$ in \eq{mu3Dsigma} is the number of quarks of a given helicity $\sigma$ in each of the target nucleons.  Equation \eqref{mu3Dsigma} is essentially \eq{mu3D} with the restriction that only partons with helicity $\sigma = \pm 1$ contribute. The generalization to the case of a nucleus having both large-$x$ quarks and gluons in its nucleons is done by the substitution
\begin{equation}
N_q^\sigma \frac{1}{2N_c} \longrightarrow \left\langle N^\sigma_R \, \frac{C_R}{K_A} \right\rangle
\end{equation}
where, again, the angle brackets denote the averaging over all the nucleons in the nucleus and $N_R^\sigma$ is the number of large-$x$ partons with color representation $R$ and with helicity $\sigma$ in each of the nucleons. Just like in the unpolarized case, the contribution in \eq{corr_sigma} comes from both currents probing the same parton (quark), which has a certain fixed helicity $\sigma$. This leads to $\delta_{\sigma \sigma'}$ on the right-hand side of \eq{corr_sigma}.

Furthermore, one can again show that correlation functions of an odd number of $j_{\sigma}^a$ are zero, while the correlation functions of an even number of $j_{\sigma}^a$ are equal to the sum of terms each of which is a product of different two-point correlators. All these results lead to the following Gaussian weight functional for the fixed-helicity currents in a given representation $r$:
\begin{align}\label{W_ab}
& \mathcal{W}^{(0)} [j_+, j_-] \propto \text{exp} \left[ - \! \int \!  \text{d}^2 x_\perp d x^- \frac{\text{tr} \left[ j_+ (x) \cdot j_{+} (x) \right]}{2\,\mu_+^2 (\un{x}, x^-) } \right] \notag \\ & \times \text{exp} \left[ - \! \int \!  \text{d}^2 x_\perp d x^- \frac{\text{tr} \left[ j_- (x) \cdot j_{-} (x) \right]}{2\,\mu_-^2 (\un{x}, x^-) } \right] ,
\end{align}
where, again, $j_{\sigma} = t^a j_{\sigma}^a$. For the unpolarized target with $\mu_+^2 = \mu_-^2 =\mu^2 /2$ and $j_+ = j_-$, the functional \eqref{W_ab} reduces to \eq{W_MV}.

Our next step is to rewrite the weight functional \eqref{W_ab} in terms of the eikonal and sub-eikonal gluon fields $\alpha (x)$ and $\beta (x)$. We continue working in Feynman gauge. Employing  equation \eqref{Box_a} with $\mu=+$ we obtain the known relation \eqref{Box_alpha} for the small-$x$ eikonal gluon field $\alpha = A^+ = a^+$,
\begin{align}\label{alpha_EOM}
\Box \alpha = - \nabla^2_\perp \alpha = j_+ + j_- = j_{\eik}^+ . 
\end{align}

Another useful relation is obtained by putting $\nu = \perp$ in \eq{g_curr},
\begin{align}\label{beta_EOM}
\Box \beta = - \nabla^2_\perp \beta = - \frac{1}{2 \langle p^+ \rangle} \nabla^2_\perp (j_+ - j_-) .
\end{align}
Here we employ the fact that $\alpha = \alpha (x^-, \un{x})$ and $\beta = \beta (x^-, \un{x})$. Inverting the system of equations \eqref{alpha_EOM} and \eqref{beta_EOM} yields
\begin{align}\label{jjab1}
\begin{pmatrix}
j_+ \\
j_-
\end{pmatrix}
=
- \thalf
\begin{pmatrix}
1 & - 2 \langle p^+ \rangle(\nabla^2_\perp)^{-1} \\
1 & + 2 \langle p^+ \rangle(\nabla^2_\perp)^{-1}
\end{pmatrix}
\cdot \nabla^2_\perp
\begin{pmatrix}
\alpha\\
\beta
\end{pmatrix}
\end{align}
with $(\nabla^2_\perp)^{-1}$ the Green function of the 2D Laplacian. The relation \eqref{jjab1} can be rewritten as 
\begin{align}\label{jjab2}
j_{\pm} = -\thalf\nabla^2_\perp \alpha \ \pm\  \langle p^+ \rangle \beta .
\end{align}

\begin{widetext}
Inserting \eq{jjab2} into \eq{W_ab}, we arrive at the following weight functional for the gluon fields:
\begin{align}\label{W_ab_final}
\mathcal{W}^{(0)} [\alpha,\beta] \propto \text{exp}\left\{ - \! \int \!  \text{d}^2 x_\perp d x^- \ \text{tr} \! \left[
(\nabla^2_\perp \alpha)^2 \, \frac{\mu_{+}^2 + \mu_{-}^2}{8\mu_{+}^2 \, \mu_{-}^2} + \left(\langle p^+ \rangle \beta\right)^2 \frac{\mu_{+}^2 + \mu_{-}^2}{2\mu_{+}^2 \, \mu_{-}^2} + (\nabla^2_\perp \alpha) \langle p^+ \rangle \beta \frac{\mu_{+}^2 - \mu_{-}^2}{2\mu_{+}^2 \, \mu_{-}^2} \right]
\right\} ,
\end{align}
\end{widetext}
where we use the abbreviated notation in which  $\mu_+^2 = \mu_+^2 (\un{x}, x^-)$, $\mu_-^2 = \mu_-^2 (\un{x}, x^-)$. 

Note that the $\beta$-field is sub-eikonal, $\beta \sim 1/\langle p^+ \rangle$, see e.g. \eq{beta_EOM}. Therefore, the combination $\langle p^+ \rangle \beta$ is of the same order in powers of energy as the $\alpha$-field. Hence, all terms in \eq{W_ab_final} are of the same order in powers of energy.


\subsection{Weight functional for the quark fields}

The weight functional for the quark fields is constructed similar to the gluon ones. We start with the two-point correlation function. As in all cases considered above, the two currents must come from the same parton. This means only the correlator of $J_\psi^{i \, \sigma}$ with $J_{\bar \psi}^{i \, \sigma} = \left( J_\psi^{i \, \sigma} \right)^\dagger$ is non-zero. A direct calculation employing Eqs.~\eqref{curr_proj6} and \eqref{curr_proj7} yields
\begin{align}\label{corr_quarks}
\langle A | J_\psi^{i \, \sigma} (\un{x}, x^-) & \, J_{\bar \psi}^{j \, \sigma'} (\un{y}, y^-) | A \rangle  = - \delta^{ij} \, \delta_{\sigma \sigma'} \, \frac{\nu_{(-\sigma)}^2 (\un{x}, x^-)}{\sqrt{2} \, \langle p^+ \rangle}  \notag   \\ & \times \, \delta (x^- - y^-) \, \delta^2 (\un{x} - \un{y}), 
\end{align}
where
\begin{align}\label{nu3Dsigma}
\nu^2_\sigma (\un{x}, x^-) = 4 \pi \as \, \frac{C_F \, N_{q}^\sigma}{N_c} \, \rho_A (\un{x}, x^-) .
\end{align}
Note that the polarization of the target quarks is opposite to the polarization of the source currents $J_\psi^{i \, \sigma}$ and $J_{\bar \psi}^{i \, \sigma}$.
Once again we assumed for simplicity that the target is made out of quarks, with $N_q^\sigma$ quarks of helicity $\sigma$ in each nucleon.
The expression \eqref{nu3Dsigma} can be easily generalized to include gluons in the target and to allow for different numbers of quarks of each polarization in the nucleons:
one needs to replace
\begin{equation}
N_q^\sigma \frac{C_F}{N_c} \longrightarrow \left\langle \eta_R \, N_R^{\sigma_R} \, \frac{C_F}{K_R} \right\rangle,
\end{equation}
in \eq{nu3Dsigma}, where $\eta_F = \eta_{\bar{F}} = 1$, $\eta_A =2$, while $N_R^{\sigma_R}$ is the number of large-$x$ partons in a given nucleon in the representation $R$ and with helicity $\sigma_R = \sigma \, \left(\delta_{R=F} + \delta_{R=\bar{F}} - \delta_{R=A} \right)$.

Just like above, the correlators of an odd number of currents $J_\psi^{i \, \sigma}$ and $J_{\bar \psi}^{i \, \sigma}$ are zero, and the correlators of an equal number of $J_\psi^{i \, \sigma}$'s and $J_{\bar \psi}^{i \, \sigma}$'s reduce to a sum of terms factorized into two-point functions \eqref{corr_quarks}. We arrive at the following Gaussian weight functional (with the color space multiplication denoted by the dots):
\begin{align}\label{Wjj}
& \mathcal{W}^{(0)} [J^\pm_\psi, J^\pm_\bpsi] \propto \text{exp} \left[ - \! \int \!  \text{d}^2 x_\perp d x^- \sqrt{2} \langle p^+ \rangle \frac{J_{\bar \psi}^+ (x) \cdot J_\psi^{+} (x)}{\nu_-^2 (\un{x}, x^-) } \right] \notag \\ & \times \text{exp} \left[ - \! \int \!  \text{d}^2 x_\perp d x^- \, \sqrt{2} \langle p^+ \rangle \, \frac{J_{\bar \psi}^- (x) \cdot J_\psi^{-} (x)}{\nu_+^2 (\un{x}, x^-) } \right] .
\end{align}
Remembering that (see \eq{Jdef}) $J_\psi^{\sigma} = \chi_\sigma^T  \, j_\psi = \chi_\sigma^T  \, [i \slashed{\pd} - m] \psi$ and employing \cite{Kovchegov:2018znm}
\begin{align}
\chi_\sigma \, \chi_\sigma^T = \frac{1}{2 \sqrt{2}} \, \gamma^0 \, \gamma^+ \, \left( 1 + \sigma \, \gamma^5 \right)
\end{align}
we rewrite \eq{Wjj} as 
\begin{widetext}
\begin{align}\label{Wppb1}
\mathcal{W}^{(0)} [\psi, {\bar \psi}] \propto \text{exp} \left\{ - \! \int \!  \text{d}^2 x_\perp d x^- \, \langle p^+ \rangle \, \left[ \left( \frac{1}{\nu_+^2} + \frac{1}{\nu_-^2} \right) {\bar \psi} \overleftarrow{\slashed{\pd}} \thalf \gamma^+ \slashed{\pd} \psi +   \left( \frac{1}{\nu_-^2} - \frac{1}{\nu_+^2} \right) {\bar \psi} \overleftarrow{\slashed{\pd}} \thalf \gamma^+ \gamma^5 \slashed{\pd} \psi \right] \right\}, 
\end{align}
where we put $m=0$ for simplicity and $\overleftarrow{\pd}$ denotes the partial derivative acting to the left. Since the quark field scales as $\psi, {\bar \psi} \sim 1/\sqrt{\langle p^+ \rangle}$, both terms in \eq{Wppb1} are of the same order in powers of energy as the terms in the exponent of \eq{W_ab_final}. Note that the fields $\psi$ and $\bar \psi$ are Grassmann variables. 

The first term in the exponent of \eq{Wppb1} is helicity-independent, while the second term projects out the helicity-dependent component of the quark field. The expression can be simplified it to 
\begin{align}\label{Wppb2}
\mathcal{W}^{(0)} [\psi, {\bar \psi}] \propto \text{exp} \left\{ \! \int \!  \text{d}^2 x_\perp d x^- \, \langle p^+ \rangle \, \left[ \left( \frac{1}{\nu_+^2} + \frac{1}{\nu_-^2} \right) {\bar \psi} \thalf \gamma^+ \nabla_\perp^2 \psi -  \left( \frac{1}{\nu_-^2} - \frac{1}{\nu_+^2} \right) {\bar \psi} \thalf \gamma^+ \gamma^5 \nabla_\perp^2 \psi  \right] \right\}, 
\end{align}
where we have remembered that transverse components dominate in $\slashed{\pd}$ and replaced $\slashed{\pd} \to - \un{\gamma} \cdot \un{\pd}$.


\subsection{Total helicity-dependent weight functional and its properties}

Combining Eqs.~\eqref{W_ab_final} and \eqref{Wppb2} we arrive at the weight functional for the helicity-generalized MV model,
\begin{align}\label{W_final}
\mathcal{W}^{(0)} [\alpha,\beta, \psi, {\bar \psi}] & \propto \text{exp}\left\{ - \! \int \!  \text{d}^2 x_\perp d x^- \ \text{tr} \! \left[
(\nabla^2_\perp \alpha)^2 \, \frac{\mu_{+}^2 + \mu_{-}^2}{8\mu_{+}^2 \, \mu_{-}^2} + \left(\langle p^+ \rangle \beta\right)^2 \frac{\mu_{+}^2 + \mu_{-}^2}{2\mu_{+}^2 \, \mu_{-}^2} + (\nabla^2_\perp \alpha) \langle p^+ \rangle \beta \frac{\mu_{+}^2 - \mu_{-}^2}{2\mu_{+}^2 \, \mu_{-}^2} \right]
\right\} \\
& \times \text{exp} \left\{  \! \int \!  \text{d}^2 x_\perp d x^- \, \langle p^+ \rangle \, \left[\frac{\nu_+^2 + \nu_-^2}{\nu_+^2 \, \nu_-^2}  \, {\bar \psi} \thalf \gamma^+ \nabla_\perp^2 \psi - \frac{\nu_+^2 - \nu_-^2}{\nu_+^2 \, \nu_-^2}  \, {\bar \psi} \thalf \gamma^+ \gamma^5 \nabla_\perp^2 \psi  \right] \right\} . \notag
\end{align}

This is the main result of this work. It should be used for flavor-singlet observables only. 

In order to make the handedness of the spinors explicit, one can use the completeness relation $1 = P_R + P_L$ of the chirality projectors $P_{R/L} = (1 \pm \gamma^5)/2$, and rewrite the second line of our main result as
\begin{equation}
\text{exp} \left\{ - \int \!  \text{d}^2 x_\perp d x^- \, \langle p^+ \rangle \, \left[\frac{1}{\nu_+^2}  \, {\bar \psi}_R \gamma^+ (i\nabla_\perp)^2 \psi_R + \frac{1}{\nu_-^2}  \, {\bar \psi}_L \gamma^+ (i\nabla_\perp)^2 \psi_L \right] \right\},
\end{equation}
where $\psi_{R/L}  = P_{R/L} \psi$ and we have included an ``$i$" with the nabla operator by analogy to the Dirac operator $i\slashed{\partial}$ and to underline the Gaussianity of the expression.
\end{widetext}

Note that the field $\beta$ needs to be scaled $\beta \to 2 \beta$ in the argument of $\mathcal{W}^{(0)}$ when the probe is a gluon and the target is made of quarks (see Appendix~\ref{App:B}). This means that, generally-speaking, our weight functional should be a matrix $\mathcal{W}^{(0)}_{PT}$ in the projectile (P)-target (T) space, with $P=q, G$, $T=q, G$. Our result \eqref{W_final} can be summarized by a relatively simple matrix in the $q, G$ space,  
\begin{align}\label{Wmatrix}
\mathcal{W}^{(0)}_{PT} [\alpha,\beta, \psi, {\bar \psi}] = \begin{pmatrix}
\mathcal{W}^{(0)} [\alpha,\beta, \psi, {\bar \psi}]  & \mathcal{W}^{(0)} [\alpha, \beta, \psi, {\bar \psi}]  \\
\mathcal{W}^{(0)} [\alpha, 2 \, \beta, \psi, {\bar \psi}]  & \mathcal{W}^{(0)} [\alpha,\beta, \psi, {\bar \psi}] 
\end{pmatrix}
.
\end{align}

In \cite{Cougoulic:2019aja}, when generalizing the JIMWLK equation to include helicity evolution, we relied on the following properties of $\mathcal{W}^{(0)}$, which were needed to justify functional integration by parts:
\begin{align}
\begin{cases}
\lim\limits_{\alpha(x) \rightarrow \pm \infty} \calw^{(0)} [\alpha,\beta, \psi, {\bar \psi}] = 0, \qquad \forall x , \\ 
\lim\limits_{\beta(x) \rightarrow \pm \infty} \calw^{(0)} [\alpha,\beta, \psi, {\bar \psi}] = 0, \qquad \forall x .
\end{cases}
\end{align}
These properties are clearly satisfied by the functional in \eq{W_final}. 

Furthermore, in \cite{Cougoulic:2019aja} we argued that the weight functional can be written as a sum of polarization independent and dependent parts,
\begin{equation}\label{ansatz0}
\calw^{(0)} = \mathcal{W}^{(0) \, unpol} + \Sigma \, \mathcal{W}^{(0) \, pol},
\end{equation}
where are $\Sigma = \pm 1$ is the target helicity (having a proton target in mind). We then conjectured that integrating out the sub-eikonal fields $\beta, \psi, {\bar \psi}$ should reduce $\calw^{(0)}$ to the MV weight functional $W^{MV}$. For the two weight functionals on the right of \eq{ansatz0} this conjecture implied the following properties:
\begin{subequations}
\begin{align}
\int \cald\beta \cald\psi \cald \bar{\psi} \ \mathcal{W}^{(0) \, unpol} [\alpha,\beta,\psi,\bpsi] &\,  = \mathcal{W}^{MV}[\alpha],  \label{W_MV2} \\
\int \cald\beta \cald\psi \cald \bar{\psi} \ \mathcal{W}^{(0) \, pol} [\alpha,\beta,\psi,\bpsi] &\, = 0 . \label{Wpol_0}
\end{align}
\end{subequations}

Writing
\begin{align}
& \mu_{+}^2 - \mu_{-}^2 = \Sigma \, |\mu_{+}^2 - \mu_{-}^2|, \\ & \nu_{+}^2 - \nu_{-}^2 = \Sigma \, |\nu_{+}^2 - \nu_{-}^2| \notag
\end{align}
we obtain from \eq{W_final}
\begin{widetext}
\begin{subequations}
\begin{align}
\mathcal{W}^{(0) \, unpol} [\alpha,\beta, \psi, {\bar \psi}] \propto & \exp \left\{ - \! \int \!  \text{d}^2 x_\perp d x^- \left[ \text{tr} \! \left[
(\nabla^2_\perp \alpha)^2 \, \frac{\mu_{+}^2 + \mu_{-}^2}{8\mu_{+}^2 \, \mu_{-}^2} + \left(\langle p^+ \rangle \beta\right)^2 \frac{\mu_{+}^2 + \mu_{-}^2}{2\mu_{+}^2 \, \mu_{-}^2} \right] - \langle p^+ \rangle \, \frac{\nu_+^2 + \nu_-^2}{\nu_+^2 \, \nu_-^2}  \, {\bar \psi} \thalf \gamma^+ \nabla_\perp^2 \psi \right] \notag
\right\} \\
& \times \cosh \left\{ - \! \int \!  \text{d}^2 x_\perp d x^- \left[ \mbox{tr} \left[ (\nabla^2_\perp \alpha) \langle p^+ \rangle \beta  \right] \frac{|\mu_{+}^2 - \mu_{-}^2|}{2\mu_{+}^2 \, \mu_{-}^2} + \langle p^+ \rangle \,  \frac{|\nu_+^2 - \nu_-^2|}{\nu_+^2 \, \nu_-^2}  \, {\bar \psi} \thalf \gamma^+ \gamma^5 \nabla_\perp^2 \psi \right] \right\} , \label{W_unpol} \\
\mathcal{W}^{(0) \, pol} [\alpha,\beta, \psi, {\bar \psi}]  \propto & \exp \left\{ - \! \int \!  \text{d}^2 x_\perp d x^- \left[ \text{tr} \! \left[
(\nabla^2_\perp \alpha)^2 \, \frac{\mu_{+}^2 + \mu_{-}^2}{8\mu_{+}^2 \, \mu_{-}^2} + \left(\langle p^+ \rangle \beta\right)^2 \frac{\mu_{+}^2 + \mu_{-}^2}{2\mu_{+}^2 \, \mu_{-}^2} \right] - \langle p^+ \rangle \, \frac{\nu_+^2 + \nu_-^2}{\nu_+^2 \, \nu_-^2}  \, {\bar \psi} \thalf \gamma^+ \nabla_\perp^2 \psi \right] 
\right\} \notag \\
& \times \sinh \left\{ - \! \int \!  \text{d}^2 x_\perp d x^- \left[ \mbox{tr} \left[ (\nabla^2_\perp \alpha) \langle p^+ \rangle \beta  \right] \frac{|\mu_{+}^2 - \mu_{-}^2|}{2\mu_{+}^2 \, \mu_{-}^2} + \langle p^+ \rangle \,  \frac{|\nu_+^2 - \nu_-^2|}{\nu_+^2 \, \nu_-^2}  \,{\bar \psi} \thalf \gamma^+ \gamma^5 \nabla_\perp^2 \psi \right] \right\} . \label{W_pol}
\end{align}
\end{subequations}
Inserting \eq{W_unpol} into the left-hand side of \eq{W_MV2} and integrating over $\beta$ yields
\begin{align}\label{Wunpol2}
& \int \cald\beta \cald\psi \cald \bar{\psi} \ \mathcal{W}^{(0) \, unpol} [\alpha,\beta,\psi,\bpsi] \propto  \exp \left\{ - \! \int \!  \text{d}^2 x_\perp d x^-  \frac{\text{tr} \! \left[
(\nabla^2_\perp \alpha)^2 \right] }{2 (\mu_{+}^2 + \mu_{-}^2)} \right\} \\ & \times \int \cald\psi \cald \bar{\psi} \ \exp \left\{ \int \!  \text{d}^2 x_\perp d x^-  \langle p^+ \rangle \, \frac{\nu_+^2 + \nu_-^2}{\nu_+^2 \, \nu_-^2}  \, {\bar \psi} \thalf \gamma^+ \nabla_\perp^2 \psi  \right\} \, \cosh  \left\{ \int \!  \text{d}^2 x_\perp d x^- \langle p^+ \rangle \, \frac{|\nu_+^2 - \nu_-^2|}{\nu_+^2 \, \nu_-^2}  \, {\bar \psi} \thalf \gamma^+ \gamma^5 \nabla_\perp^2 \psi  \right\}. \notag
\end{align}
The $\psi$ and $\bar \psi$ integrals in \eq{Wunpol2} just affect the normalization factor, which we do not keep track of here. We arrive at
\begin{align}\label{Wunpol3}
 \int \cald\beta \cald\psi \cald \bar{\psi} \ \mathcal{W}^{(0) \, unpol} [\alpha,\beta,\psi,\bpsi]  \propto  \exp \left\{ - \! \int \!  \text{d}^2 x_\perp d x^-  \frac{\text{tr} \! \left[
(\nabla^2_\perp \alpha)^2 \right] }{2 (\mu_{+}^2 + \mu_{-}^2)} \right\}  .
\end{align}
Finally, noticing that $\mu_{+}^2 + \mu_{-}^2 = \mu^2$ (since the total number of large-$x$ partons is the sum of the numbers of spin-up and spin-down partons) we recognize that the right-hand side of \eq{Wunpol3} is identical to that of \eq{WMVa}. We have thus verified \eq{W_MV2}. 

Similarly, inserting \eq{W_pol} into the left-hand side of  \eq{Wpol_0} and integrating over $\beta$ we arrive at
\begin{align}\label{Wpol3}
& \int \cald\beta \cald\psi \cald \bar{\psi} \ \mathcal{W}^{(0) \, pol} [\alpha,\beta,\psi,\bpsi] \propto  \exp \left\{ - \! \int \!  \text{d}^2 x_\perp d x^-  \frac{\text{tr} \! \left[
(\nabla^2_\perp \alpha)^2 \right] }{2 (\mu_{+}^2 + \mu_{-}^2)} \right\} \\ & \times \int \cald\psi \cald \bar{\psi} \ \exp \left\{ \int \!  \text{d}^2 x_\perp d x^-  \langle p^+ \rangle \, \frac{\nu_+^2 + \nu_-^2}{\nu_+^2 \, \nu_-^2}  \, {\bar \psi} \thalf \gamma^+ \nabla_\perp^2 \psi  \right\} \, \sinh  \left\{ \int \!  \text{d}^2 x_\perp d x^- \langle p^+ \rangle \, \frac{|\nu_+^2 - \nu_-^2|}{\nu_+^2 \, \nu_-^2}  \, {\bar \psi} \thalf \gamma^+ \gamma^5 \nabla_\perp^2 \psi  \right\} =0. \notag
\end{align}
\end{widetext}
The $\psi$ and $\bar \psi$ integrals in \eq{Wpol3} give zero. To see this one can notice that the integrand of the $\psi$ and $\bar \psi$ integrals simply changes sign under a combination of parity and time reversal (PT), since the argument of the hyperbolic sine changes sign under PT, while the exponent remains the same. Parity transformation can be thought of as an integration variable change in that integral, since it interchanges $\psi_L \stackrel{\mbox{P}}{\longleftrightarrow} \psi_R$ with $\psi_{R,L} = \thalf (1 \pm \gamma^5) \psi$. The integration measure is invariant under parity since it can be written as
\begin{align}
\int \cald\psi \cald \bar{\psi} \propto \int \cald\psi_L \cald \bar{\psi}_L \cald\psi_R \cald \bar{\psi}_R
\end{align}
where the proportionality is valid up to a multiplicative constant. Time reversal $\psi \stackrel{\mbox{T}}{\longrightarrow} - \gamma^1 \, \gamma^3 \, \psi$ also preserves the functional integration measure since $\det [\gamma^1 \, \gamma^3] =1$. Hence the PT transformation recasts the last line of \eq{Wpol3} as a negative of itself: therefore, the integral is zero. (Note that, under PT, $(x^-, \un{x}) \stackrel{\mbox{PT}}{\longrightarrow} - (x^-, \un{x})$, $\mu_+^2\stackrel{\mbox{PT}}{\longleftrightarrow}\mu_-^2$, and $\nu_+^2\stackrel{\mbox{PT}}{\longleftrightarrow}\nu_-^2$: none of these transformations affects the integrals in \eq{Wpol3}.) We have thus demonstrated that the property \eqref{Wpol_0} is also satisfied by our weight functional \eqref{W_final}.

Finally, in \cite{Cougoulic:2019aja} we have argued that the weight functionals should satisfy the following properties:
\begin{subequations}\label{WVconds}
\begin{align}
& \hspace*{-2mm} \int \!\! \cald\alpha \cald\beta \cald\psi \cald \bar{\psi} \mathcal{W}^{(0) \, unpol} [\alpha,\beta,\psi,\bpsi] V_\xvec^{pol} (\alpha,\beta,\psi,\bpsi)  = 0 , \label{WVpol} \\
&  \hspace*{-2mm} \int \cald\alpha \cald\beta \cald\psi \cald \bar{\psi} \ \mathcal{W}^{(0) \, pol} [\alpha,\beta,\psi,\bpsi]  \, V_\xvec (\alpha) =0. \label{WpolV}
\end{align}
\end{subequations} 
Here $V_\xvec (\alpha) \equiv V_\xvec [+\infty, - \infty]$ is the light-cone infinite Wilson line in the projectile direction, which depends only on the eikonal gluon field $\alpha$. This implies that the condition \eqref{WpolV} follows from the condition \eqref{Wpol_0} which we have just verified.

The ``polarized Wilson line" $V_\xvec^{pol} (\alpha,\beta,\psi,\bpsi)$ is defined above in \eq{eq:Wpol_all}. It consists of eikonal light-cone Wilson lines with sub-eikonal operator insertions. There are two terms with different sub-eikonal insertions in $V_\xvec^{pol} (\alpha,\beta,\psi,\bpsi)$: one terms is linear in $\beta$ and  
another one is linear in ${\bar \psi} \thalf \gamma^+ \gamma^5 \psi$. Using $\mathcal{W}^{(0) \, unpol}$ from \eq{W_unpol} one can show that the condition \eqref{WVpol} is satisfied by using the PT-symmetry argument to factor out integrals over $\psi$ and $\bar \psi$ and then integrate over $\beta$ to obtain zero in the $\sim \beta$ term. The ${\bar \psi} \thalf \gamma^+ \gamma^5 \psi$ term is zero after $\beta$-integration if we again invoke the PT-symmetry for the fermion integrals. We conclude that both conditions in \eqref{WVconds} are also satisfied by our weight functional \eqref{W_final}.

In this derivation, we focused on the leading contribution in power of $A^{1/3}$. It would be possible to extend the present derivation to higher (suppressed by powers of $A^{-1/3}$) orders by following higher-order extension of the MV model (see \cite{Jeon:2005cf} and references therein) applied to sub-eikonal fields.


\section{Conclusions}
\label{conc}

In this work we have generalized the MV model to describe helicity-dependent flavor-singlet observables, in addition to the standard unpolarized quantities included in the original MV model. The resulting helicity-MV weight functional in $\pd_\mu A^\mu =0$ gauge is given by \eq{W_final}. One can show that the weight functional \eqref{W_final} is also valid in $A^-=0$ gauge. It can be used to find the expectation values of helicity-dependent (or independent) operators using the standard MV prescription,
\begin{equation}
\label{EQ:Operator_average_helicity}
\langle \hcalo_{pol} \rangle = \int \cald\alpha \cald\beta \cald\psi \cald \bar{\psi} \ \hcalo_{pol} \ {\cal W}^{(0)} [\alpha,\beta,\psi,\bar{\psi}].
\end{equation}
These expectation values will be valid in the quasi-classical limit of QCD, which neglects small-$x$ evolution. The latter evolution can be included with the help of the helicity generalization of JIMWLK equation, which was derived in \cite{Cougoulic:2019aja} and is given by \eq{EQ:Helicity_evolution_equation} above. Together with our result \eqref{W_final}, the helicity-JIMWLK evolution \eqref{EQ:Helicity_evolution_equation} allows one to describe any helicity-dependent flavor-singlet observable at small $x$. In the large-$N_c$ limit this description for helicity PDFs and for transverse momentum dependent helicity PDFs has been done in \cite{Kovchegov:2015pbl,Kovchegov:2016zex,Kovchegov:2016weo,Kovchegov:2017jxc,Kovchegov:2017lsr,Kovchegov:2018znm} with the analysis of the large-$N_c \& N_f$ limit recently completed in \cite{Kovchegov:2020hgb}. The formalism based on Eqs.~\eqref{W_final} and \eqref{EQ:Helicity_evolution_equation} will allow for generalization of those results to any phenomenologically relevant value of $N_c$ and $N_f$.


\section*{Acknowledgments}

We would like to thank Andrey Tarasov and Raju Venugopalan for discussions.

This material is based upon work supported by The U.S. Department of
Energy, Office of Science, Office of Nuclear Physics under Award
Number DE-SC0004286.



\appendix

\section{Diagrammatic check: the $\beta$-field}
\label{App:A}

In this Appendix, we argue that in Feynman gauge, the class of diagrams contributing to $\beta=F_{12}$ of a single parton, and, through Eqs.~\eqref{eq:Wpol_all} and \eqref{M:UpolFull}, to the corresponding parts of Born-level scattering amplitudes can be written in a compact form as 
\begin{equation}\label{Answer}
ig^2  \lambda \delta_{\lambda \lambda'} \Lambda \delta_{\Lambda \Lambda'} \times \eta_{R_P}  \left(t_{R_P}^e \otimes t_{R_T}^e\right) 
\end{equation}
with $R_P$ the irrep of the probe and $R_T$ the irrep of the target, $\lambda/\lambda'$ and $\Lambda/\Lambda'$ the helicities of the projectile and target before/after the scattering, while $\eta_F = \eta_{\bar{F}} = 1$, $\eta_A =2$. Equation \eqref{Answer} contains the helicity-dependent part of the $2 \to 2$ Born scattering amplitudes for (anti)quarks and gluons mediated by the $\beta$-field. It serves as a verification of the scaling in $\eta_{R_P}$ expected from the definitions of polarized Wilson lines $V^{pol,g}$ and $U^{pol,g}$ (defined as the $F^{12}$-parts of Eqs.~\eqref{eq:Wpol_all} and \eqref{M:UpolFull}, respectively), i.e., that $W^{(R_P)pol,g} \sim \eta_{R_P}$ with $W^{(F_P)}=V$ and $W^{(A_P)} = U$ for the fundamental ($F_P$) and adjoint ($A_P$) representations of the projectile. \eq{Answer} also cross-checks $\eta_{R_T}$-independence of the sub-eikonal source current \eqref{j_subeik_gluon} found in the main text.

Note that $\beta=F_{12}$ also contains a term quadratic in the gluon field. Below we will also discuss the quadratic term $[A_1,A_2] \subset F_{12}$ and its diagrammatic content at the sub-eikonal level.


\subsection{Kinematics}

We will have to consider several scattering processes $ab \rightarrow ab$ in this Subsection. To keep track of them, we will always indicate by the first particle $a$ the probe and by the second particle $b$ the target. The corresponding helicity-dependent (and, hence, sub-eikonal) amplitude will be denoted $\mathcal{M}^{ab\rightarrow ab}$. (While, indeed, $2 \to 2$ scattering is completely target--projectile symmetric, the individual terms in the polarized Wilson lines \eqref{eq:Wpol_all} and \eqref{M:UpolFull} are not, and the target--projectile symmetry is restored only when all such terms are included.) 

We work in the Feynman gauge ($\partial_\mu A^\mu = 0$) and consider massless partons. We chose a frame where the projectile is light-cone minus moving with an initial momentum $q$ and zero transverse momentum $\qvec= \zv$ and helicity $\sigma$ for a quark (resp. $\lambda$ for a gluon.) The target is light-cone plus moving with an initial momentum $p$, zero transverse momentum $\pvec = \zv$, and helicity $\Sigma$ for a quark (resp. $\Lambda$ for a gluon.) The momentum exchanged is denoted by $k$ such that we have the momenta $q+k$ and $p-k$ in the final state for the projectile and the target, respectively.

Since we are only interested in the helicity-dependent part of the amplitude, we have the following ``helicity trick" for the gluon polarization vectors,
\begin{subequations}
\begin{equation}
\label{EQ:HelicityTrick}
\left[ \underline{\epsilon}_{\lambda} \right]_\alpha  \left[ \underline{\epsilon}^*_{\lambda'} \right]_{\beta} \supset \frac{-i\lambda \delta_{\lambda\lambda'}}{2} \epsilon_{\alpha\beta}^\perp,
\end{equation}
which defines a substitution needed to extract the helicity-dependent ($\sim \lambda \delta_{\lambda\lambda'}$) part in the product of two polarization vectors (with $\epsilon_{\alpha\beta}^\perp$ the two-dimensional Levi-Civita symbol). Using \eq{EQ:HelicityTrick} we define the following two substitutions (for the minus moving gluons and for the plus moving gluons):
\begin{align}
\label{EQ:HelicityTrick1}
\left[ \epsilon_{\lambda}(q)\right]_\alpha \left[ \epsilon^*_{\lambda'}(q+k)\right]_\beta \supset 
\frac{-i\lambda \delta_{\lambda \lambda'}}{2}  \epsilon_{\alpha\beta}^\perp , \\
\label{EQ:HelicityTrick2}
\left[ \epsilon_{\Lambda}(p)\right]_\chi \left[ \epsilon^*_{\Lambda'}(p-k)\right]_\rho \supset \frac{-i\Lambda \delta_{\Lambda \Lambda'}}{2}   \epsilon_{\chi \rho}^\perp,
\end{align} 
\end{subequations}
where we have neglected the energy-suppressed terms proportional to $\epsilon_{\lambda/\Lambda}^{\pm}$.


\subsection{Gluon target}

First we consider a polarized gluon target. 

\paragraph{Gluon-gluon scattering.}
In the kinematics specified above we have
\begin{center}
\begin{tikzpicture}
\draw[gluon] (0,1) node[left] {$q,\lambda,\alpha, a$} -- (2,.5) -- (4,1) node[right] {$q+k,\lambda',\beta, b$};
\draw[gluon] (0,0) node[left] {$p,\Lambda,\chi, c$} -- (2,.5) -- (4,0) node[right] {$p-k,\Lambda',\rho, d$};
\draw[fill=black!20!white] (2,.5) ellipse (.7 and .4);
\end{tikzpicture}
\end{center}
where we explicitly label the momenta flowing from left to right, helicities, colors, and Lorentz indices contracting with the polarization vectors.
As we will soon see, the sum of all diagrams will be proportional in color space to the $t$-channel diagram color tensor. Thus we will consider color-stripped amplitude for simplicity, denoted by $m$ with a subscript indicating the corresponding channel.

Consider the $t$-channel diagram. The color-stripped amplitude reads 
\begin{align}
i\,m_t =& \frac{-i(-ig)^2}{t} \left[ \epsilon_{\lambda}(q)\right]_\alpha \left[ \epsilon^*_{\lambda'}(q+k)\right]_\beta \left[ \epsilon_{\Lambda}(p)\right]_\chi \left[ \epsilon^*_{\Lambda'}(p-k)\right]_\rho
\nonumber \\
& \times \left\{ 2k^\alpha g^{\beta\mu} - 2k^\beta g^{\alpha\mu} - (2q+k)^\mu g^{\alpha\beta}\right\} g_{\mu\nu} \nonumber \\
& \times  \left\{ 2k^\chi g^{\nu\rho} -2k^\rho g^{\chi\nu} + (2p-k)^\nu g^{\chi\rho} \right\}  .
\end{align}
The helicity-dependent part of $m_t$ is obtained by using Eqs.~\eqref{EQ:HelicityTrick1} and \eqref{EQ:HelicityTrick2}, which yields
\begin{align}\label{mtGG}
i\,m_t \! \supset& \frac{-i(-ig)^2}{t} \frac{-i\lambda \delta_{\lambda \lambda'}}{2} \frac{-i\Lambda \delta_{\Lambda \Lambda'}}{2}
\epsilon_{\alpha\beta}^\perp  \epsilon_{\chi \rho}^\perp g_{\mu\nu} 
4 k^{[\alpha} g^{\beta]\mu}  k^{[\chi} g^{\rho]\nu} \notag \\
=& \, ig^2 \frac{-i\lambda \delta_{\lambda \lambda'}}{2} \frac{-i\Lambda \delta_{\Lambda \Lambda'}}{2} \times 16 ,
\end{align}
where the square brackets $[\alpha \ldots \beta ]$ denote commutators of indices, $[\alpha \ldots \beta ] = \alpha \ldots \beta - \beta \ldots \alpha$. For the reasons that will be evident latter on, the result in \eq{mtGG} can be conveniently re-written as
\begin{equation}\label{MtGG}
i\,\mathcal{M}_t^{gg\rightarrow gg} \supset ig^2 \left[ 
\raisebox{-5pt}{
\begin{tikzpicture}
\draw[gluon] (0,0) -- ++(1,0);
\draw[gluon] (0,.5) -- ++(1,0);
\draw[gluon] (.5,0) -- ++(0,.5);
\end{tikzpicture}}
\right]\frac{-i\lambda \delta_{\lambda \lambda'}}{2} \frac{-i\Lambda \delta_{\Lambda \Lambda'}}{2} \times 4 \, \eta_{A_P} \eta_{A_T},
\end{equation}
where we have introduced back the color tensor
\begin{equation}
\label{Eq:ColorTchan}
\raisebox{-15pt}{
\begin{tikzpicture}[scale=.9]
\draw[gluon] (-1,.5) node[left] {$a$} -- (0,.4) -- (1,.5) node[right] {$b$};
\draw[gluon] (0,.4) -- (0,-.4);
\draw[gluon] (-1,-.5) node[left] {$c$} -- (0,-.4) -- (1,-.5) node[right] {$d$};
\end{tikzpicture}} = if^{aeb} if^{cde}.
\end{equation}

\smallbreak
The $s$-channel color-stripped amplitude $m_s$ reads
\begin{align}
i\, & m_s  \! = \! \frac{-i(-ig)^2}{s}  \! \left[ \epsilon_{\lambda}(q)\right]_\alpha \! \left[ \epsilon^*_{\lambda'}(q+k)\right]_\beta  \!\left[ \epsilon_{\Lambda}(p)\right]_\chi  \! \left[ \epsilon^*_{\Lambda'}(p-k)\right]_\rho  \notag \\
\times & \left\{ 2p^\alpha g^{\chi\mu} - 2q^\chi  g^{\alpha\mu} +(q-p)^\mu g^{\alpha\chi} \right\} g_{\mu\nu}  \\
 \times & \left\{ (2p+2q)^\beta g^{\rho\nu} - (2p+2q)^\rho g^{\beta\nu} + (2k-p+q)^\nu g^{\beta\rho}  \right\} . \notag
\end{align}
The helicity-dependent part of $m_s$ is again obtained by using Eqs.~\eqref{EQ:HelicityTrick1} and \eqref{EQ:HelicityTrick2},
\begin{align}\label{ms100}
i\,m_s \supset& \frac{-i(-ig)^2}{s} \frac{-i\lambda \delta_{\lambda \lambda'}}{2} \frac{-i\Lambda \delta_{\Lambda \Lambda'}}{2}
\epsilon_{\alpha\beta}^\perp  \epsilon_{\chi \rho}^\perp g^{\alpha\chi} g^{\beta\rho} \times (-s) \nonumber \\
=& \, ig^2 \frac{-i\lambda \delta_{\lambda \lambda'}}{2} \frac{-i\Lambda \delta_{\Lambda \Lambda'}}{2} \times (-2) ,
\end{align}
where the color tensor is
\begin{equation}
\label{Eq:ColorSchan}
\raisebox{-15pt}{
\begin{tikzpicture}[scale=.9]
\draw[gluon] (-1,.5) node[left] {$a$} -- (-.4,0);
\draw[gluon] (-1,-.5) node[left] {$c$} -- (-.4,0);
\draw[gluon] (-.4,0) -- (.4,0);
\draw[gluon] (.4,0) -- (1,.5) node[right] {$b$};
\draw[gluon] (.4,0) -- (1,-.5) node[right] {$d$};
\end{tikzpicture}}
= if^{ace} if^{dbe}.
\end{equation}

The helicity-dependent part of the color-stripped amplitude for the $u$-channel, $m_u$, is obtained similarly to the $s$-channel
\begin{equation}
i\,m_u = \, ig^2 \frac{-i\lambda \delta_{\lambda \lambda'}}{2} \frac{-i\Lambda \delta_{\Lambda \Lambda'}}{2} \times (+2),
\end{equation}
where the relative sign simply follows from the high-energy limit where $u \approx -s$ for $s \gg |t|$.
Its color tensor is
\begin{equation}
\label{Eq:ColorUchan}
\raisebox{-15pt}{
\begin{tikzpicture}[scale=.9]
\draw[gluon] (-1,.5) node[left] {$a$} to[out=0,in=90] (.4,0);
\draw[gluon] (-1,-.5) node[left] {$c$} -- (-.4,0);
\draw[gluon] (-.4,0) -- (.4,0);
\draw[gluon] (-.4,0) to[out=90,in=180] (1,.5) node[right] {$b$};
\draw[gluon] (.4,0) -- (1,-.5) node[right] {$d$};
\end{tikzpicture}}
= if^{bce} if^{dae}.
\end{equation}

Summing both $s$- and $u$-channels with the use of Jacobi identity
\begin{equation}
\label{Eq:Jacobi}
\raisebox{-12pt}{
\begin{tikzpicture}[scale=.9]
\draw[gluon] (-1,.5) -- (0,.4);
\draw[gluon,Red] (0,.4) -- (1,.5);
\draw[gluon] (0,.4) -- (0,-.4);
\draw[gluon] (0,-.4) -- (-1,-.5);
\draw[gluon] (0,-.4) -- (1,-.5);
\end{tikzpicture}}
-
\raisebox{-12pt}{
\begin{tikzpicture}[scale=.9]
\draw[gluon] (-1,.5) -- (-.4,0);
\draw[gluon] (-1,-.5) -- (-.4,0);
\draw[gluon] (-.4,0) -- (.4,0);
\draw[gluon,Red] (.4,0) -- (1,.5);
\draw[gluon] (.4,0) -- (1,-.5);
\end{tikzpicture}}
+
\raisebox{-12pt}{
\begin{tikzpicture}[scale=.9]
\draw[gluon] (-1,.5)  to[out=0,in=90] (.4,0);
\draw[gluon] (-1,-.5) -- (-.4,0);
\draw[gluon] (-.4,0) -- (.4,0);
\draw[gluon,Red] (-.4,0) to[out=90,in=180] (1,.5);
\draw[gluon] (.4,0) -- (1,-.5);
\end{tikzpicture}}
=0,
\end{equation}
allows us to write (now including the color factor, denoted by the birdtrack diagram)
\begin{equation}\label{MsuGG}
i\,\mathcal{M}_{s+u}^{gg\rightarrow gg} \supset ig^2 \left[ 
\raisebox{-5pt}{
\begin{tikzpicture}
\draw[gluon] (0,0) -- ++(1,0);
\draw[gluon] (0,.5) -- ++(1,0);
\draw[gluon] (.5,0) -- ++(0,.5);
\end{tikzpicture}}
\right]\frac{-i\lambda \delta_{\lambda \lambda'}}{2} \frac{-i\Lambda \delta_{\Lambda \Lambda'}}{2} \times (-2) .
\end{equation}

In addition to the $s$, $t$, and $u$ channel diagrams, the four gluon vertex diagram also contributes to the gluon-gluon scattering. It contains helicity-dependent sub-eikonal terms. The four gluon vertex comes with three different color tensors, given by Eqs.~\eqref{Eq:ColorTchan}, \eqref{Eq:ColorSchan}, and \eqref{Eq:ColorUchan}.
The color-stripped amplitudes associated with the $t$-, $s$-, and $u$-channel color tensors are
\begin{subequations}
\begin{align}
i\,m_4^t \supset& \ ig^2 \frac{-i\lambda \delta_{\lambda \lambda'}}{2} \frac{-i\Lambda \delta_{\Lambda \Lambda'}}{2} \times (- 4),\\
i\,m_4^s \supset& \ ig^2 \frac{-i\lambda \delta_{\lambda \lambda'}}{2} \frac{-i\Lambda \delta_{\Lambda \Lambda'}}{2} \times (-2), \\
i\,m_4^u \supset& \ ig^2 \frac{-i\lambda \delta_{\lambda \lambda'}}{2} \frac{-i\Lambda \delta_{\Lambda \Lambda'}}{2} \times (+2).
\end{align}
\end{subequations}
Adding those, and using the Jacobi identity \eqref{Eq:Jacobi}, one finds the net contribution of the 4-gluon vertex diagram,
\begin{align}\label{M4GG}
i\,\mathcal{M}_4^{gg\rightarrow gg} \supset& \ ig^2 \left[ 
\raisebox{-5pt}{
\begin{tikzpicture}
\draw[gluon] (0,0) -- ++(1,0);
\draw[gluon] (0,.5) -- ++(1,0);
\draw[gluon] (.5,0) -- ++(0,.5);
\end{tikzpicture}}
\right]\frac{-i\lambda \delta_{\lambda \lambda'}}{2} \frac{-i\Lambda \delta_{\Lambda \Lambda'}}{2} \times (- 6) .
\end{align}

Combining Eqs.~\eqref{MtGG}, \eqref{MsuGG}, and \eqref{M4GG} one finds
\begin{equation}\label{MGG1}
i\,\mathcal{M}^{gg\rightarrow gg} \supset ig^2 \left[ 
\raisebox{-5pt}{
\begin{tikzpicture}
\draw[gluon] (0,0) -- ++(1,0);
\draw[gluon] (0,.5) -- ++(1,0);
\draw[gluon] (.5,0) -- ++(0,.5);
\end{tikzpicture}}
\right]\frac{-i\lambda \delta_{\lambda \lambda'}}{2} \frac{-i\Lambda \delta_{\Lambda \Lambda'}}{2} \times 8 .
\end{equation}
Note that $\mathcal{M}^{gg\rightarrow gg}_t : (\mathcal{M}_{s+u}^{gg\rightarrow gg} + \mathcal{M}_4^{gg\rightarrow gg}) = 2: (-1)$. This will be useful below. 

The overall color tensor \eqref{Eq:ColorTchan} can also be written as
\begin{equation}
\raisebox{-5pt}{
\begin{tikzpicture}
\draw[gluon] (0,0) -- ++(1,0);
\draw[gluon] (0,.5) -- ++(1,0);
\draw[gluon] (.5,0) -- ++(0,.5);
\end{tikzpicture}}
= - t_{A_P}^e \otimes t_{A_T}^e
\end{equation}
where $(t_A^e)_{ba} = - i f^{bae}$, $t^e_{A_P}$ and $t^e_{A_T}$ are the adjoint color factors of the projectile and the target, written with the $(t^e_A)_{\text{final}, \text{initial}}$ ordering of indices for the incoming and outgoing projectile or target gluons.
It is now convenient to identify the numerical factor in \eq{MGG1} with 
\begin{equation}\label{Eq:scaling_GG}
8 = 4 \times \eta_{A_P} \times (\eta_{A_P} -1) 
= 4 \times \eta_{A_P} .
\end{equation}
This yields
\begin{align}
\label{Eq:Mgg_result}
i\,\mathcal{M}^{gg\rightarrow gg} \supset ig^2  \lambda \delta_{\lambda \lambda'} \Lambda \delta_{\Lambda \Lambda'} \times \eta_{A_P}  \left(t_{A_P}^e \otimes t_{A_T}^e\right) .
\end{align}


\smallbreak
\paragraph{Quark-gluon scattering.}
Now, we consider the process with a minus moving quark and a plus moving gluon, i.e., we are changing the probe representation $A \rightarrow F$ compared to the $gg \rightarrow gg$ case. The process is illustrated by
\begin{center}
\begin{tikzpicture}
\draw[thick] (0,1) node[left] {$q,\sigma$} -- (2,.5) -- (4,1) node[right] {$q+k,\sigma'$};
\draw[gluon] (0,0) node[left] {$p,\Lambda,\chi, a$} -- (2,.5) -- (4,0) node[right] {$p-k,\Lambda',\rho, b$};
\draw[fill=black!20!white] (2,.5) ellipse (.7 and .4);
\end{tikzpicture}
\end{center}
with the same kinematics as specified above.  Factoring the color tensor 
\begin{equation}
\label{Eq:ColorTchanQG}
if^{abc}t^c
\end{equation}
out of the amplitude, we obtain the $t$-channel color-stripped amplitude 
\begin{align}
i\,m_t =& -\frac{ig^2}{t}\bar{u}_{\sigma'}(q+k) \gamma^\mu u_\sigma(q) \ g_{\mu\nu} \ \left[ \epsilon_\Lambda(p) \right]_\chi \left[ \epsilon_{\Lambda'}^*(p-k)\right]_{\rho} \nonumber \\
&\times \left[ 2k^\chi g^{\rho\nu} -2k^\rho g^{\chi\nu} +(2p-k)^\nu g^{\chi\rho}\right].
\end{align}
For a light-cone minus-moving quark, the helicity-dependent part of the matrix element is obtained using
\begin{align}
\label{Eq:qline}
\bar{u}_{\sigma'}(q+k) \gamma^\mu u_\sigma(q) \supset& \frac{-i\sigma \delta_{\sigma \sigma'}}{2} \times (-2) \epsilon^{\mu\gamma}_\perp k_\gamma .
\end{align}
In addition, using \req{EQ:HelicityTrick2}, one finds
\begin{align}
i\,m_t \supset& -\frac{ig^2}{t} \frac{-i\sigma \delta_{\sigma \sigma'}}{2}  \frac{-i\Lambda \delta_{\Lambda \Lambda'}}{2} 
(-8) k^\gamma \epsilon_{\mu\gamma}^\perp g^{\mu\rho}  \epsilon_{\chi\rho}^\perp k^{\chi} \nonumber \\
=&  -ig^2 \frac{-i\sigma \delta_{\sigma \sigma'}}{2}  \frac{-i\Lambda \delta_{\Lambda \Lambda'}}{2} \times 8 .
\end{align}
The numerical factor can be written as 
\begin{equation}
8= 4\times \eta_{F_P} \times \eta_{A_T} ,
\end{equation}
which can be compared with the factor $4\times \eta_{A_P} \times \eta_{A_T}$ in $\mathcal{M}_t^{gg\rightarrow gg}$.

Next we factor out the color tensors 
\begin{equation}
\label{Eq:ColorSchanQG}
t^bt^a \qquad \text{and} \qquad t^a t^b
\end{equation}
from the $s$- and $u$-channel amplitudes. The $s$-channel color-stripped amplitude reads
\begin{align}\label{Eq:ims_qg0}
i\,m_s = & \, - \frac{i g^2}{s} \bar{u}_{\sigma'}(q+k) \gamma^\rho (\slashed{p}+\slashed{q}) \gamma^\chi u_\sigma(q) \notag \\ & \times \left[ \epsilon_\Lambda(q) \right]_\chi \left[ \epsilon_{\Lambda'}^*(q+k)\right]_{\rho} .
\end{align}
Using \req{EQ:HelicityTrick2}, it yields
\begin{align}
i\,m_s \supset& -ig^2 \frac{-i\sigma \delta_{\sigma \sigma'}}{2}  \frac{-i\Lambda \delta_{\Lambda \Lambda'}}{2} \times (-4).
\end{align}
The $u$-channel yields a similar result
\begin{equation}
i\,m_u \supset -ig^2 \frac{-i\sigma \delta_{\sigma \sigma'}}{2}  \frac{-i\Lambda \delta_{\Lambda \Lambda'}}{2} \times (+4).
\end{equation}

Using $[t^a, t^b] = i f^{abc}t^c$, the full amplitude reads
\begin{equation}\label{Mqg1}
i\,\mathcal{M}^{qg \rightarrow qg} \supset -ig^2 
\left[ 
if^{abc}t^c
\right]
\frac{-i\sigma \delta_{\sigma \sigma'}}{2} \frac{-i\Lambda \delta_{\Lambda \Lambda'}}{2} \times 4
\end{equation}
where it will be convenient to identify the numerical factor as 
\begin{equation}\label{Eq:scaling_QG}
4 = 4 \times \eta_{F_P} \times (\eta_{A_T} -1) = 4\times \eta_{F_P}.
\end{equation}
Note that, similar to the gluon-gluon scattering, $\mathcal{M}^{qg\rightarrow qg}_t : \mathcal{M}_{s+u}^{qg\rightarrow qg} = 2: (-1)$.

Since $if^{abc} = (t^c_{A})_{ba}$, \eq{Mqg1} finally yields
\begin{align}\label{Eq:Result_QG}
i\,\mathcal{M}^{qg\rightarrow qg} \supset ig^2  \Lambda \delta_{\Lambda \Lambda'} \sigma \delta_{\sigma \sigma'} 
\times \eta_{F_P}  \left(t_{F_P}^e \otimes t_{A_T}^e \right).
\end{align}
This is exactly \req{Eq:Mgg_result} with the substitution $\eta_{A_P} \to \eta_{F_P}$, $t_{A_P}^e \to t_{F_P}^e$ accounting for the gluon probe being replaced by a quark probe.


\subsection{Quark target}

In the case of a quark target, only the $t$-channel diagrams belong to the $\beta$-field. The $s$- and $u$-channel contributions to the helicity-dependent amplitude will be discussed in Appendix \ref{App:B}, as they arise from the quark field dependent terms in Eqs.~\eqref{eq:Wpol_all} and \eqref{M:UpolFull}. 


\paragraph{Gluon-quark scattering.}
In the above-given kinematics, we consider the scattering process with a minus-moving gluon and a plus-moving quark:
\begin{center}
\begin{tikzpicture}
\draw[gluon] (0,1) node[left] {$q,\lambda,\alpha, a$} -- (2,.5) -- (4,1) node[right] {$q+k,\lambda',\beta, b$};
\draw[thick] (0,0) node[left] {$p,\Sigma$} -- (2,.5) -- (4,0) node[right] {$p-k,\Sigma'$};
\draw[fill=black!20!white] (2,.5) ellipse (.7 and .4);
\end{tikzpicture}
.
\end{center}
Compared to the gluon-gluon scattering case, we are changing the representation of the target while keeping the representation of the probe.

Factoring the color tensor 
\begin{equation}
\label{Eq:ColorTchanGQ}
if^{bac}t^c
\end{equation}
out of the amplitude $\mathcal{M}_t$, the color-stripped amplitude reads
\begin{align}
i\,m_t =& -\frac{ig^2}{t}\bar{u}_{\Sigma'}(p-k) \gamma^\mu u_\Sigma(p) \ g_{\mu\nu} \ \left[ \epsilon_\lambda(q) \epsilon_{\lambda'}^*(q+k)\right]_{\alpha\beta} \nonumber \\
&\times \left[ 2k^\alpha g^{\beta\nu} -2k^\beta g^{\alpha\nu} -(2q+k)^\nu g^{\alpha\beta}\right] .
\end{align}
In order to pull out the helicity-dependent sub-eikonal contribution we use
\begin{equation}
\label{Eq:pline}
\bar{u}_{\Sigma'}(p-k) \gamma^\mu u_\Sigma(p) \supset \frac{-i \Sigma \delta_{\Sigma\Sigma'}}{2} \times 2\epsilon^{\mu\gamma}_\perp k_\gamma,
\end{equation}
in addition to \req{EQ:HelicityTrick1} to find
\begin{align}
i\,m_t \supset ig^2 \frac{-i \lambda \delta_{\lambda\lambda'}}{2}\frac{-i \Sigma \delta_{\Sigma\Sigma'}}{2} \times 8 .
\end{align}
Notice that this factor also reads
\begin{equation}\label{Eq:scaling_GQ}
8 = 4 \times \eta_{A_P} \times \eta_{F_T} = 4 \times \eta_{A_P}
\end{equation}
such that we get
\begin{align}\label{Eq:Mt_GQ}
i\,\mathcal{M}_t^{gq\rightarrow gq} \supset ig^2  \lambda \delta_{\lambda \lambda'} \Sigma \delta_{\Sigma \Sigma'} \times \eta_{A_P}  \left(t_{A_P}^e \otimes t_{F_T}^e\right) .
\end{align}


\smallbreak
\paragraph{Quark-quark scattering.}
Using Eqs.~\eqref{Eq:pline} and \eqref{Eq:qline}, it is straightforward to find the $t$-channel exchange amplitude, 
\begin{align}\label{Eq:Mt_QQ}
i\,\mathcal{M}_t^{qq \rightarrow qq} \supset ig^2 \sigma \delta_{\sigma \sigma'} \Sigma \delta_{\Sigma \Sigma'}
\times \eta_{F_P}  \left( t_{F_P}^e \otimes t_{F_T}^e \right),
\end{align}
which can also be obtained from \req{Eq:Mt_GQ} by the substitution $\eta_{A_P} \rightarrow \eta_{F_P}$, $t_{A_P}^e \to t_{F_P}^e$,  corresponding to the replacement of a gluon probe by a quark probe.


\subsection{Discussion}

The overall factor $\eta_{R_P}$ in the scattering amplitudes \eqref{Eq:Mgg_result}, \eqref{Eq:Result_QG}, \eqref{Eq:Mt_GQ}, and \eqref{Eq:Mt_QQ} above, depending on whether the probe is a quark ($R_P = F_P$) or a gluon ($R_P = A_P$), is expected from the definition of the polarized Wilson lines in Eqs.~\eqref{eq:Wpol_all} and \eqref{M:UpolFull}. The $F^{12}$-dependent parts of the polarized Wilson lines \eqref{eq:Wpol_all} and \eqref{M:UpolFull},  which we denote summarily by $W^{(R_P) pol,g}$, can be written as
\begin{align}
\label{EQ:Def_of_W_pol_g}
& W^{(R_P) pol,g}_\xvec = \eta_{R_P} \frac{ig p_1^+}{s}\\
& \times \int\limits_{-\infty}^\infty \text{d}x^-\, W^{(R_P)}_\xvec [\infty , x^-] \, \beta(x^-,\xvec) \, W^{(R_P)}_\xvec [x^- , -\infty ], \nonumber
\end{align}
with the factor of $\eta_{R_P} =1$ for $R_P=F_P$ and $\eta_{R_P} =2$ for $R_P=A_P$ accounting for the factor of 2 difference between the $F^{12}$-terms in Eqs.~\eqref{M:UpolFull} and \eqref{eq:Wpol_all}. (In the kinematics of this Appendix $p_1^+ = p^+$.) Our above calculation confirms the $\eta_{R_P}$-dependence of the operator in \eq{EQ:Def_of_W_pol_g}.

We have shown here that the $t$-channel diagrams are not the only ones to contain sub-eikonal and helicity-dependent contributions to the scattering amplitudes, but that $s$-, or $u$-channel diagrams also do contribute. This is a feature of the helicity-dependent part of the scattering amplitude and it should be contrasted with the usual helicity-independent part of the amplitude which, in the eikonal approximation, is only given by the $t$-channel gluon exchange.
Fortunately, in the Feynman gauge it is possible to absorb those new contributions in the sub-eikonal source current \eqref{EQ:Source_subeik_gluon},
\begin{align*}
j^{a \, \mu}_{subeik}(x^-, {\underline x}) \simeq 
- g \frac{\epsilon_\perp^{\mu i}\partial^i_\perp}{2 \langle p^+ \rangle} \!\!\!
\sum_{\substack{\sigma = \pm \\ R_T=\{q,\bar{q},g\}}} \!\!\!\!\!
\sigma \,  \hat{\ag}^\dagger_{R_T, x,\sigma} \cdot t^a_{R_T} \cdot \hat{\ag}_{R_T, x,\sigma} ,
\end{align*}
which is independent of the representation of the target in the sense of being independent of $\eta_{R_T}$.

It is useful to keep in mind that the target wave function averages of the amplitudes discussed here vanish due to color algebra. This is not a surprise since the averages are proportional to $\langle \beta \rangle = 0$, which is zero due to color algebra. The average is only non-vanishing when there is at least one $\alpha$-field, for example $\langle \beta \alpha \rangle \neq 0$.


\subsection{The quadratic term in the $\beta$ field} 

Identification of the $s$-, $u$-channel, and $4$-gluon vertex diagrams in $gg \rightarrow gg$ scattering as contributing to the $\beta$-field may be surprising. The same applies to identifying the $s$- and $u$-channel diagrams in  $qg \rightarrow qg$ scattering as contributing to the $\beta$-field. Below we will argue this point in more details.

Consider a polarized gluon target. Using the notation of \eq{Aexp} we separate the target gluon field into the $A_0^\mu$ incoming/outgoing plane wave and the order-$g$ $t$-channel gluon field $a^\mu$. By its definition, the $\beta$-field is
\begin{align}\label{bf12}
\beta \equiv f^{12} = \partial^1 a^{2} - \partial^2 a^{1} - i  g \, [ A_0^{1},  A_0^{2} ] . 
\end{align}
The $a^\mu$ field is illustrated in middle panel of the top row of \fig{FIG:Subeik_sources}: it can be straightforwardly identified as the field responsible for the $t$-channel gluon exchange in the $gg \rightarrow gg$ and $qg \rightarrow qg$ scattering considered above in this Appendix. 

To identify the remaining $- i  g \, [ A_0^{1},  A_0^{2} ]$ term in \eq{bf12}, we first consider \eq{g_curr} re-written as 
\begin{align}\label{g_curr2}
\partial_\mu f^{\mu\nu}  = & \,  \Box a^\nu - i g \, \partial_\mu [ A_0^{\mu},  A_0^{\nu} ] \notag \\  = & \, i g \, [ A_{0,\mu}, \partial^\mu A^{\nu}_0] - i g \, [ A_{0,\mu}, \partial^\nu A_0^{\mu} ] ,
\end{align}
and note that only the first term in the second line of \eq{g_curr2} gives a helicity-dependent contribution. Indeed, the second term contains $A_{0,\mu} \ldots A_0^{\mu}$, leading to a scalar product of the two polarization vectors contained in $A_0$ fields: such a scalar product cannot give us the helicity dependence $\sim \Lambda \delta_{\Lambda \Lambda'}$ that we need (cf. \eq{EQ:HelicityTrick2}). 

We then compare \eq{g_curr2} to \eq{Box_a}, re-written as (neglecting the order-$g^2$ term)
\begin{align}\label{Box_a2}
\Box a^{\nu} = 2 i g \, [A_{0,\mu}, \pd^\mu A_0^{\nu}] - i g \, [A_{0,\mu}, \pd^\nu A_0^{\mu}],
\end{align}
where, again, only the first term on the right is helicity-dependent, and is twice larger than the first term in the second line of \eq{g_curr2}. We conclude that the effect of the $- i g \, \partial_\mu [ A_0^{\mu},  A_0^{\nu} ]$ term in the first line of \eq{g_curr2} is to reduce $\Box a^\nu$ in half (as long as helicity-dependent contribution is concerned). More specifically, concentrating on helicity-dependent terms only, and noticing that in Feynman gauge only transverse components of $A_0^\mu$ contribute due to the high-energy approximation for the polarization vector, we put $\nu = 2$ in  \eq{g_curr2} and re-write it as
\begin{align}\label{g_curr3}
\partial_1 \beta  =  \Box a^2 - i g \, \partial_1 [ A_0^{1},  A_0^{2} ] = & \, i g \, [ A_{0,i}, \partial^i A^{2}_0] ,
\end{align}
while \eq{Box_a2} becomes
\begin{align}\label{Box_a3}
\Box a^{2} = 2 i g \, [A_{0,i}, \pd^i A_0^{2}] = 2 \, \partial_1 \beta , 
\end{align}
where $i = 1, 2$. We arrive at
\begin{align}
- i  g \, [ A_0^{1},  A_0^{2} ] = - \beta, 
\end{align}
such that \eq{g_curr3} can be cast as
\begin{align}
\beta = 2 \beta - \beta. 
\end{align}
This means, the ratio of the $t$-channel helicity-dependent contribution due to $a^\mu$ to the remaining contributions coming from $- i  g \, [ A_0^{1},  A_0^{2} ]$ is $2:(-1)$. Above we saw that $\mathcal{M}^{gg\rightarrow gg}_t : (\mathcal{M}_{s+u}^{gg\rightarrow gg} + \mathcal{M}_4^{gg\rightarrow gg}) = 2: (-1)$ and $\mathcal{M}^{qg\rightarrow qg}_t : \mathcal{M}_{s+u}^{qg\rightarrow qg} = 2: (-1)$. We, therefore, identify the contributions of the $s$-, $u$-channel and 4-gluon vertex diagrams in the $gg \rightarrow gg$ scattering with $g \, f^{abc} \, A_0^{1, b} \, A_0^{2, c} \subset f^{a, 12} = \beta^a$ and the $s$- and $u$-channel diagrams in the $qg \rightarrow qg$ scattering with $- i  g \, [ A_0^{1},  A_0^{2} ] \subset f^{12} = \beta$. Diagrammatically this identification is illustrated by the left pair of arrows in \fig{FIG:Subeik_int_vertices}.  

To further illustrate how the $s$- and $u$-channel diagrams, with two vertices coupling to two $A_0$ fields at different space-time locations as shown in lines two and four of \fig{FIG:Subeik_int_vertices}, can contribute to a local operator like $g \, f^{abc} \, A_0^{1, b} (x) \, A_0^{2, c} (x) $ in $\beta^a$, let us consider the $s$-channel diagram for $gg \rightarrow gg$. As explicitly calculated above, the $1/s$ factor arising from the propagator of the $s$-channel diagram is canceled by the same factor in the numerator, resulting in the $s$-independent expression \eqref{ms100}. Since the propagator is thus canceled, the two vertices coupling to $A_0$ fields are now at the same location. The same argument can be applied to the $u$-channel diagram for $gg \rightarrow gg$. Another way of saying this is that the helicity-dependent parts of the $s$- and $u$-channel diagrams in $gg \rightarrow gg$ scattering have the same $s$- and $u$-dependence as the 4-gluon vertex diagram (all of them are constant). The four-gluon vertex is explicitly local, resulting in the two $A_0$ fields being at the same location for the $s$-, $u$- and 4-gluon channels. 

The same argument can be applied to the $s$- and $u$-channel diagrams in the $qg \rightarrow qg$ scattering, resulting on both $A_0$ fields being probed locally. Alternatively, consider the $s$-channel diagram for $qg \rightarrow qg$ evaluated in \req{Eq:ims_qg0} (see also the second row of \fig{FIG:Subeik_int_vertices}). In order to produce a sub-eikonal contribution, one has to pick from the propagator's numerator the term $\gamma^- p^+ \subset \slashed{p} + \slashed{q}$. Replacing $p^+ \to \pd_-$ one can rewrite \req{Eq:ims_qg0} in the following form: 
\begin{align}\label{ms200}
im_s &\propto \int_{-\infty}^{\infty} dx^- \int_{-\infty}^{x^-} dy^- A^i_0(x^-, \un{x}) \, \frac{\partial}{\partial y^-} A^j_0(y^-, \un{x}) \notag \\
&\rightarrow \int_{-\infty}^{\infty} dx^- A^i_0(x^-, \un{x}) \, A^j_0(x^-, \un{x}),
\end{align}
where $i,j = 1,2$. 
This is an explicitly local operator. Finally, the link to $f_{12}$ in particular is explicit when considering the helicity trick \req{EQ:HelicityTrick2} which will pick the contribution $\epsilon_{\mu\nu}^\perp [A_0^\mu A_0^\nu](x) \subset f_{12}(x)$. Alternatively, adding the $u$-channel  $qg \rightarrow qg$ diagram to \eq{ms200} would convert $A^i_0(x) \, A^j_0(x) \to [A^i_0(x) , A^j_0(x)] \subset f_{12} (x)$.

We have thus shown that, for a gluon target, the sum of scattering diagrams in all channels, as shown in \fig{FIG:Subeik_int_vertices}, is accounted for by a single local operator $F^a_{12} (x)$ (for the helicity-dependent sub-eikonal contribution only). 

For the quark target the analogous conclusion is easy to obtain: since there is no $A_0$-field in the quark target, only the $t$-channel gluon exchange, as illustrated in the top left panel of \fig{FIG:Subeik_sources}, contributes to the $\beta$-field. However, for the case of the quark target we have the $\psi_0$ field (see \eq{psi0}), which is the incoming/outgoing plane-wave analogue of $A_0$. The contribution of this field to the polarized Wilson lines \eqref{eq:Wpol_all} and \eqref{M:UpolFull} and to the corresponding diagrams shown in \fig{FIG:Subeik_int_vertices} is analyzed separately in Appendix~\ref{App:B}.


\section{Treatment of the $\psi_0$ and $\bpsi_0$ fields}

\label{App:B}

The goal of this Appendix is to analyze the $s$- and $u$-channel diagram contributions to the quark field $\psi$-dependent parts of the polarized Wilson lines \eqref{eq:Wpol_all} and \eqref{M:UpolFull} applied to scattering on a polarized quark target (with the $t$-channel contributions discussed in detail in the main text). We argue that the $s$- and $u$-channel diagrams result in the bilinear local quark operator shown in Figs.~\ref{FIG:Subeik_int_vertices} (right-most diagrams in lines two and four) and \ref{FIG:Subeik_sources} (right-most diagram in the lower line). They are accounted for in the helicity-extended MV model constructed in this work by the modification of the weight functional for the gluon probe scattering on the target quark, as done in \eq{Wmatrix}. 


\subsection{Operator analysis}

The solution of the equation of motion \eqref{Eq:Psi_EoM} for $\psi$ can be written as (cf. \eq{psi0})
\begin{equation}
\psi = \psi_0 + \slashed{\Delta} j_\psi,
\end{equation}
with $\psi_0$ the solution of the homogeneous equation and $j_\psi$ defined in \req{Jpsi}. The plane-wave field $\psi_0$ of the incoming/outgoing (anti-)quarks will contribute to the polarized Wilson lines \eqref{eq:Wpol_all} and \eqref{M:UpolFull} through the $\psi$-dependent terms in those operators. Since the incoming ($\psi_0$) and outgoing ($\bpsi_0$) fields have to be in the same nucleon in the nucleus, it is natural to expect that the two fields have to be at the same space-time position, similar to the gluon field $A_0$ in Appendix~\ref{App:A}. This is indeed the case: we note that both Eqs.~\eqref{eq:Wpol_all} and \eqref{M:UpolFull} contain the following operator, which we can simplify as
\begin{align}\label{local_op}
& p_1^+ \int\limits_{-\infty}^\infty dx_1^- \int \limits_{x_1^-}^\infty dx_2^- \ \bpsi_0^i(x_2^-,\xvec)\gamma^+\gamma^5 \psi_0^j(x_1^-,\xvec) \notag \\ & = i \int\limits_{-\infty}^\infty dx_1^- \int \limits_{x_1^-}^\infty dx_2^- \ \bpsi_0^i (x_2^-,\xvec)\gamma^+\gamma^5 \pd^+ \psi_0^j (x_1^-,\xvec) \notag \\ & = i \int\limits_{-\infty}^\infty dx^- \ \bpsi_0^i (x^-,\xvec)\gamma^+\gamma^5 \psi_0^j (x^-,\xvec) ,
\end{align}
obtaining a local operator with fundamental color indices $i,j$. The contribution of the local operator \eqref{local_op} to the polarized Wilson lines is shown in lines two and four of \fig{FIG:Subeik_int_vertices}, where the right set of arrows illustrates the separation of the $\psi$-dependent parts of these operators into the local and non-local terms. (The latter couple to the $\slashed{\Delta} j_\psi$ part of the field $\psi$.)

In arriving at \eq{local_op}  we have neglected the non-trivial contributions of the Wilson lines in the $\psi$-dependent terms in Eqs.~\eqref{eq:Wpol_all} and \eqref{M:UpolFull} connecting points $x_1^-$ and $x_2^-$, since such interaction is negligible inside a single nucleon. In general, to derive the interaction with the quark target coming from these $\psi$-dependent terms, one has to substitute the decompositions \eqref{Aexp} and \eqref{psi0} into the $\psi$-dependent parts of Eqs.~\eqref{eq:Wpol_all} and \eqref{M:UpolFull}, take $j_\psi = - g \slashed{A_0} \psi_0$ such that $\slashed{\Delta} j_\psi$ is order-$g$, and expand the resulting operators to order-$g^4$, while keeping in mind that the source is a single quark in a nucleon. Since, in Feynman gauge, only transverse components of the $A_0^\mu$ field contribute, the regular Wilson lines in the $\psi$-dependent terms of \eqref{eq:Wpol_all} and \eqref{M:UpolFull} will only couple to $a^+ = \alpha$. In the end one will be left with two terms, a Dirac bilinear in $\psi_0$ shown in \eq{local_op} (and accompanied by one power of $a^+ = \alpha$), and a Dirac bilinear in $\slashed{\Delta} j_\psi$, whose expectation value can be found using the weight functional \eqref{Wppb2}. (In addition, one has to note that the flavor-singlet contribution is proportional to $V^{pol}_\xvec + V^{pol \, \dagger}_\xvec$, instead of just $V^{pol}_\xvec$ from \eq{eq:Wpol_all}: this means only cut diagrams contribute \cite{Kovchegov:2018znm}.) The results of this operator expansion can also be reproduced diagrammatically while employing the fact that $s$- and $u$-channel $2 \to 2$ scattering diagrams where quark and gluon interchange their roles as the projectile and target, that is, $s$- and $u$-channels for $qg \leftrightarrow gq$, are sub-sub-eikonal and can be neglected in our sub-eikonal calculation. 

At the diagrammatic level, the result \eqref{local_op} corresponds to the interaction of the probe with quark target through the $s$- and/or $u$-channel diagrams, with the former shown in lines two and four of \fig{FIG:Subeik_int_vertices}. These diagrams indeed are helicity-dependent and sub-eikonal, and can also be recast from a bi-local operator in $\psi_0$ to the local operator \eqref{local_op}, following the same steps as shown above (cf. \eq{ms200} and the discussion around it).

Note that the color indices in \eq{local_op} are not contracted, and can be projected onto the color-singlet and color-octet states. Below we will show that, for the flavor-singlet observables considered in this work, the color-singlet projection of \eqref{local_op} vanishes, leaving only the color-octet one, that is, an axial current operator 
\begin{align}\label{axial_beta}
\bpsi_0 \gamma^+\gamma^5 t^a \psi_0 = \frac{1}{g} \left[ j^a_+ - j^a_- \right] = \frac{1}{g} \langle 2 p^+ \rangle \, \beta^a ,
\end{align}
where in the last step we have used \eq{beta_EOM} for the $\beta$-field sources by the same quark $\psi_0$. We see that it is natural to absorb the contribution of the local operator in \eq{local_op} into the $\beta$-field. One possibility is to \textit{``move"} the local part \eqref{axial_beta} of the operator bilinear in $\psi$ in Eqs.~\eqref{eq:Wpol_all} and \eqref{M:UpolFull} to $W^{(R_P) pol,g}$. Alternatively, and we follow this approach in the main text, one can account for the term \eqref{axial_beta} by modifying the weight functional, but, as we will show below, only for the gluon probe scattering on the quark target (see \eq{Wmatrix}). This underlines a novel property of sub-eikonal and helicity-dependent fields in the helicity-dependent MV model: they are described by a probe-dependent weight functional.


\subsection{Diagrams}

As mentioned before, in this Appendix it is understood that the target is a polarized quark. We will now analyze the $s$- and $u$-channel diagrams for $gq \to gq$ and $q/\bar{q} + q \to q/\bar{q} + q$ in the flavor-singlet channel, with the aim to assess the contribution of the operator \eqref{local_op}. 

\paragraph{Gluon probe.}
Employing parity (or, equivalently, projectile--target duality), it is easy to recover that the helicity-dependent sum of the $s$- and $u$-channel diagrams is 
\begin{equation}\label{Eq:gq_s_u_chan}
\, \mathcal{M}^{gq \rightarrow gq}_{s+u} \supset
- ig^2  \lambda \delta_{\lambda \lambda'} \Sigma \delta_{\Sigma \Sigma'} \times  \left(t_{A_1}^e \otimes t_{F_2}^e\right).
\end{equation}
To obtain \eq{Eq:gq_s_u_chan} one can simply use \eq{Eq:Mt_GQ} and notice that the contribution of the $s$- and $u$-channels is (-1/2) of the $t$-channel, as we observed above for the helicity-dependent part. Indeed, the sum of \req{Eq:gq_s_u_chan} with \req{Eq:Mt_GQ} is (modulo an interchange of the color generators)
\begin{equation}
\mathcal{M}^{gq \rightarrow gq}_{s+u+t} = \mathcal{M}^{qg \rightarrow qg} = \thalf \mathcal{M}_t^{gq \rightarrow gq},
\end{equation}
where the $\mathcal{M}^{qg \rightarrow qg}$ is given by \req{Eq:Result_QG} while $\mathcal{M}_t^{gq \rightarrow gq}$ is given by \eq{Eq:Mt_GQ}. We see that adding the $s$- and $u$-channels reduces the contribution \eqref{Eq:Mt_GQ} of the polarized quark target to the $\beta$-field by a factor of 2. Note that this conclusion is made for the gluon projectile. Therefore, for $gq \to gq$ scattering, we can take into account the contributions of the $s$- and $u$-channels either by replacing
\begin{equation}
\beta \rightarrow \thalf \beta
\end{equation}  
in \eq{M:UpolFull} (with $F_{12}^a = \beta^a$) or, equivalently, by rescaling
\begin{align}
\beta \to 2 \beta
\end{align}
in the $Gq$ weight functional $\mathcal{W}^{(0)}_{Gq}$ in \eq{Wmatrix}. 


\smallbreak
\paragraph{Quark and anti-quark probes.}
Since we are only interested in the flavor-singlet case, one has to sum the contributions of the quark and anti-quark scattering on the polarized quark probe. For a fixed configuration of the target nucleus, the flavor content in the target is fixed. To be specific, let us denote by $f$ the flavor of the target quark that we scatter on in a nucleon. The $s$-channel (resp. $u$-channel) diagram exists only when the flavor of the probe is $\bar{f}$ (resp. $f$). This is to be contrasted with the $t$-channel diagrams presented in Appendix~\ref{App:A}, which exist for all flavors of the probe.
In order to obtain the helicity-dependent contribution from the $s$-channel diagrams, one can use the Fierz transformation and pick the axial-vector contribution,
\begin{align}
&\bar{v}_\sigma(q) \gamma^\mu u_\Sigma(p) \times \bar{u}_{\Sigma'}(p-k) \gamma_\mu v_{\sigma'}(q+k) = \notag \\
&\thalf\thalf \bar{v}_\sigma(q) \gamma^\mu \gamma^5 v_{\sigma'}(q+k) \times \bar{u}_{\Sigma'} \gamma_\mu \gamma^5 u_\Sigma(p) + \cdots ,
\end{align}
where the ellipsis indicates either further energy-suppressed or helicity-independent contributions.
This approach yields the color-stripped amplitude
\begin{equation}\label{Eq:imsqq}
i\, m_s \supset ig^2 \sigma\delta_{\sigma\sigma'} \Sigma \delta_{\Sigma\Sigma'} \times (-1).
\end{equation}
Similar calculation for the color-stripped $u$-channel diagram gives
\begin{equation}\label{Eq:imuqq}
i\, m_u \supset ig^2 \sigma\delta_{\sigma\sigma'} \Sigma \delta_{\Sigma\Sigma'} \times (+1).
\end{equation}
The relative sign between the contributions in Eqs.~\eqref{Eq:imsqq} and \eqref{Eq:imuqq} can be understood as coming from the definition \eqref{Eq:antispinnors_def_b}, the high-energy limit where $u \sim -s$, and the sign change in the amplitude under the interchange of two fermions.

Let us recast associated color tensor in term of $t$-channel irreps. Using the completeness relation, the $s$-channel color tensor decomposes into
\begin{equation}\label{Eq:csqq}
\raisebox{-5pt}{
\begin{tikzpicture}
\draw[thick] (0,.25) -- (.25,0) -- (0,-.25);
\draw[thick,-<] (0,.25) -- ++(0.125,-0.125);
\draw[gluon] (.25,0) -- (.75,0);
\draw[thick] (1,.25) -- (.75,0) -- (1,-.25);
\draw[thick,->] (1,.25) -- ++(-0.125,-0.125);
\end{tikzpicture}}
=\frac{C_F}{N_c} 
\raisebox{-5pt}{
\begin{tikzpicture}
\draw[thick] (0,.25) -- (1,.25);
\draw[thick,-<] (0,.25) -- ++(0.5,0);
\draw[thick] (0,-.25) -- (1,-.25);
\draw[thick,->] (0,-.25) -- ++(0.5,0);
\end{tikzpicture}}
-\frac{1}{N_c}
\raisebox{-5pt}{
\begin{tikzpicture}
\draw[thick] (0,.25) -- (1,.25);
\draw[thick,-<] (0,.25) -- ++(0.3,0);
\draw[thick] (0,-.25) -- (1,-.25);
\draw[thick,->] (0,-.25) -- ++(0.3,0);
\draw[gluon] (.5,.25) -- (.5,-.25);
\end{tikzpicture}}
\end{equation}
and the $u$-channel tensor into
\begin{equation}\label{Eq:cuqq}
\raisebox{-5pt}{
\begin{tikzpicture}
\draw[thick] (0,.25) -- (.25,.25) to[out=-90,in=180]  (.75,-.25) -- (1,-.25);
\draw[thick,->] (0,.25) -- ++(.2,0);
\draw[thick] (1,.25) -- (.75,.25) to[out=-90,in=0]  (.25,-.25) -- (0,-.25);
\draw[gluon] (.25,.25) -- (.75,.25);
\draw[thick,-<] (1,.25) -- ++(-.2,0);
\end{tikzpicture}}
=\frac{C_F}{N_c} 
\raisebox{-5pt}{
\begin{tikzpicture}
\draw[thick] (0,.25) -- (1,.25);
\draw[thick,->] (0,.25) -- ++(0.5,0);
\draw[thick] (0,-.25) -- (1,-.25);
\draw[thick,->] (0,-.25) -- ++(0.5,0);
\end{tikzpicture}}
+\frac{1}{N_c}
\raisebox{-5pt}{
\begin{tikzpicture}
\draw[thick] (0,.25) -- (1,.25);
\draw[thick,->] (0,.25) -- ++(0.3,0);
\draw[thick] (0,-.25) -- (1,-.25);
\draw[thick,->] (0,-.25) -- ++(0.3,0);
\draw[gluon] (.5,.25) -- (.5,-.25);
\end{tikzpicture}}
\ \ .
\end{equation}
This is new compared to the $gq \rightarrow gq$ case: in addition to the color-octet channel both the $s$- and $u$-channel diagrams contain a color-singlet contribution, albeit valued in different vector-spaces, namely in $\bar{F} \otimes F \rightarrow \bar{F} \otimes F$ for the $s$ channel and in $F \otimes F \rightarrow F \otimes F$ for the $u$ channel.


It is now straightforward to see that the expectation value of the local part of the operator $\langle[\bpsi_0^i \gamma^+ \gamma^5 \psi_0^j ](x)\rangle$ (see \eq{local_op}), in the flavor-singlet case, vanishes: the color-octet terms in Eqs.~\eqref{Eq:csqq} and \eqref{Eq:cuqq} are zero due to the color algebra alone, while the color-singlet terms in \eqref{Eq:csqq} and \eqref{Eq:cuqq} cancel each other due to the sign difference in Eqs.~\eqref{Eq:imsqq} and \eqref{Eq:imuqq}. (In calculating flavor-singlet observables one has to add the contributions of the quark and anti-quark projectile scattering on the same target, $\sim \left[ V^{pol}_\xvec + V^{pol \, \dagger}_\xvec \right]$, \cite{Kovchegov:2018znm}.)

Next consider the flavor-singlet correlation function $\sum_f\ \langle V^{pol}_\xvec + V^{pol \, \dagger}_\xvec \rangle$, concentrating on the contribution of the local operator \eqref{local_op}. Expanding in the powers of $\alpha$, and employing the fact that  $\langle[\bpsi_0^i \gamma^+ \gamma^5 \psi_0^j ](x)\rangle =0$, we get
\begin{align}\label{flsing}
& \sum_f\ \langle V^{pol}_\xvec + V^{pol \, \dagger}_\xvec \rangle \supset \\
& \sum_f\ \left\langle V_\xvec [+\infty, x^-] [\bpsi_0 \gamma^+ \gamma^5 \psi_0 ](x) V_\xvec [x^-, -\infty] \right. \notag \\ & \left. + V^\dagger_\xvec [+ \infty, x^-] [\bpsi_0 \gamma^+ \gamma^5 \psi_0 ](x) V^\dagger_\xvec [x^-, -\infty] \right\rangle = {\cal{O}} (\alpha^2). \notag 
\end{align}
(We suppress the color factors and indices for simplicity.) We see that the operator \eqref{flsing} is zero at the order-$g^4$, which is the order at which we are interested in an interaction with a single nucleon in the quasi-classical approximation at hand.  
Diagrammatically the cancellation can be pictured as follows (the thin vertical (red) line indicates the final state cut),
\begin{equation}\label{sing_cancel}
\raisebox{-10pt}{
\begin{tikzpicture}
\draw[thick] (0,.25) -- (.25,0) -- (0,-.25);
\draw[thick,-<] (0,.25) -- ++(0.125,-0.125);
\draw[gluon] (.25,0) -- (.75,0);
\draw[thick] (1,.25) -- (.75,0) -- (1,-.25);
\draw[thick,->] (1,.25) -- ++(-0.125,-0.125);

\draw[thick] (1,.25) -- (2,.25);
\draw[thick] (1,-.25) -- (2,-.25);
\draw[gluon] (1.5,.25) -- (1.5,-.25);

\draw[Red] (1.1,.4) -- (1.1,-.4);
\end{tikzpicture}}
\ + \  
\raisebox{-10pt}{
\begin{tikzpicture}
\draw[thick] (0,.25) -- (.25,.25) to[out=-90,in=180]  (.75,-.25) -- (1,-.25);
\draw[thick,->] (0,.25) -- ++(.2,0);
\draw[thick] (1,.25) -- (.75,.25) to[out=-90,in=0]  (.25,-.25) -- (0,-.25);
\draw[gluon] (.25,.25) -- (.75,.25);
\draw[thick,-<] (1,.25) -- ++(-.2,0);

\draw[thick] (1,.25) -- (2,.25);
\draw[thick] (1,-.25) -- (2,-.25);
\draw[gluon] (1.5,.25) -- (1.5,-.25);

\draw[Red] (1.1,.4) -- (1.1,-.4);
\end{tikzpicture}}
\ =\ 0.
\end{equation}
This cancellation only occurs in the flavor-singlet case, and is a consequence of the relative sign between a) Eqs.~\eqref{Eq:imsqq} and \eqref{Eq:imuqq}, b) the color-octet contribution in Eqs.~\eqref{Eq:csqq} and \eqref{Eq:cuqq}, and c) the sign difference in the coupling of the gluon to the quark and the anti-quark to the right of the cut. We conclude that the $s$- and $u$-channel diagrams do not contribute to the $q/\bar{q} + q \to q/\bar{q} + q$ scattering in the flavor-singlet channel, and need not to be included into our formalism. 

It is worth mentioning that in the flavor non-singlet case the cancellation \eqref{sing_cancel} no longer applies, since one would need to subtract rather than add the $qq \to qq$ and $\bar{q}q \to \bar{q}q$ contributions \cite{Kovchegov:2018znm}. We expect that, similar to the $s$- and $u$-channel diagrams in $gq \to gq$ case considered above, the resulting contribution can be absorbed into the $\beta$-field, or into the corresponding weight functional. The $\beta$-field would then be flavor-dependent, since the $s$- and $u$-channel diagrams may not exist for every probe of a given flavor scattering on a particular target. Further investigation of the flavor non-singlet channel, beyond the large-$N_c$ limit studied in \cite{Kovchegov:2016zex}, is left for future work.


\providecommand{\href}[2]{#2}\begingroup\raggedright\endgroup

\end{document}